\documentclass[aps,prd,superscriptaddress,nofootinbib,11pt]{revtex4}
\usepackage[english]{babel}
\usepackage[utf8]{inputenc}
\usepackage{graphicx}   
\usepackage{slashed}
\usepackage{epstopdf}
\usepackage{verbatim}   
\usepackage{color}      
\usepackage{subfigure}  
\usepackage{multirow}
\usepackage{hyperref}   
\usepackage{float}
\usepackage{epsfig,rotating}
\usepackage{amsmath,amssymb}
\usepackage{dsfont}
\usepackage{slashed}
\restylefloat{table}
\numberwithin{equation}{section}

\newcommand{\vx}{\vec{x}}
\newcommand{\vp}{\vec{p}}

\newcommand{\vk}{\vec{k}}

\newcommand{\be}{\begin{equation}}
\newcommand{\ee}{\end{equation}}
\newcommand{\bea}{\begin{eqnarray}}
\newcommand{\eea}{\end{eqnarray}}

\newcommand{\ket}[1]{|#1\rangle}
\newcommand{\bra}[1]{\langle#1|}

\begin{document}
\title{Probing dynamical axion quasiparticles with two-photon correlations.}

\author{Daniel Boyanovsky}
\email{boyan@pitt.edu} \affiliation{Department of Physics, University of Pittsburgh, Pittsburgh, PA 15260}

 \date{\today}

\begin{abstract}
Dynamical axion (quasi) particles are emergent collective excitations in topological magnetic insulators or in Weyl semimetals that break parity and time reversal invariance  . They couple to electromagnetism via a topological Chern-Simons term, leading to their decay into two photons. We extend the Weisskopf-Wigner formulation of atomic spontaneous emission to the   quantum field theory of dynamical axion quasiparticles, allowing us to obtain the quantum two-photon state emerging from axion decay in real time. This state features \emph{hyperentanglement} in  momentum and polarization with a distinct polarization pattern, a consequence of the parity and time reversal breaking of the axion-photon interaction. Polarization aspects of this two-photon state are studied by introducing quantum Stokes operators. Whereas the two-photon quantum state features  vanishing \emph{averages} of the degree of polarization and polarization asymmetry, there are non-trivial momentum correlations of the Stokes operators. In particular momentum correlations of the \emph{polarization asymmetry} can be obtained directly from   coincident momentum and polarization resolved  two photon detection. Correlations of Stokes operators are directly related to momentum and polarization resolved Hanbury-Brown Twiss second order coherences. This relationship  suggests      two-photon correlations as a direct probe of dynamical axion quasiparticles.   Similarities and differences with parametrically down converted photons and other systems where spontaneous emission yield hyperentangled two photon states are recognized, suggesting experimental avenues similar to tests of Bell inequalities to probe dynamical axion quasiparticles with coincident two photon detection.

\end{abstract}

\keywords{}

\maketitle

\section{Introduction}\label{sec:intro}

The axion is a hypothetical pseudoscalar particle that was originally  introduced in Quantum Chromodynamics (QCD) as a solution of the strong CP problem\cite{PQ,weinaxion,wil} and is   a potentially viable    dark matter candidate\cite{marsh,sikivie1,press,abbott,fischler}.    An important feature that characterizes the axion field $\phi(\vx,t)$  is  its   coupling to electromagnetism   via a topological Chern-Simons term $ \propto \epsilon_{\mu\nu\rho\sigma}F^{\mu\nu} F^{\rho \sigma} \propto \vec{E}\cdot\vec{B}$,  giving rise to novel phenomena, in other words,    axion electrodynamics\cite{wilczekaxion}. While the particle physics axion has not yet been unambiguously observed, remarkably, in condensed matter physics dynamical axion quasiparticles emerge as collective excitations in magnetic topological insulators that break time reversal invariance\cite{xiao,rundong,jing,nomura,narang}, charge density waves in Weyl semimetals with broken parity and time reversal invariance\cite{gooth,gos,yu,mottola} or in multilayered metamaterials\cite{wilczekshapo}. These collective excitations  couple to electromagnetism via a Chern-Simons term, just as the cosmological axion, leading to  topological magnetoelectric effects such as Faraday and Kerr rotations\cite{liang,tse,ahn}, with potential impact in  the detection of the particle physics axion\cite{ishi,jan}. In presence of a constant magnetic field dynamical axions couple \emph{linearly} to the electric field, resulting in a hybridization of photons and axions, namely \emph{axionic polaritons}\cite{rundong,zhu}. This axion-photon mixing is also at the heart of the Primakoff effect that is proposed as an experimental method of detection of the dark matter axion\cite{marsh2,chig}. Observation of  emergent axionic collective excitations in condensed matter systems not only may lead to novel platforms for quantum information based on topological excitations, but may also provide a bridge towards a complementary fundamental understanding   and  detection of its particle physics relative\cite{jan,marsh2,chig}. Furthermore,  emergent dynamical axion quasiparticles mix and hybridize\cite{boyhybrid} with the dark matter axion via a common two-photon channel as a consequence of their topological coupling to electromagnetism. This fundamental aspect  may also yield   a detection mechanism of dark matter axions by harnessing the condensed matter dynamical axion excitations.

 In ref.\cite{zaletel} it  has been proposed that the vanishing of the bulk gap at the surface of magnetic topological insulators yield an enhancement of the fluctuations of the dynamical axion field near the surface providing a possible platform for their detection. The dynamical and dark matter axions are both manifest as coherent oscillations of the axion field.  Recently the observation of such coherent oscillation   induced by an antiferromagnetic magnon   has been reported in two dimensional ($MnBi_2Te_4$)\cite{jianq} and interpreted as direct evidence for a dynamical axion quasiparticle. This observation  bolsters the case for a deeper understanding of axion quasiparticles in three dimensions with possible extrapolation to their dark matter relatives in particle physics.

\vspace{1mm}

\textbf{Motivation and main  objectives:}
The coupling of the dynamical axion,   (quasi)particle associated with a pseudoscalar field $\delta \theta(\vx)$,  to the electromagnetic field $\propto \delta \theta(\vx)\vec{E}(\vx)\cdot\vec{B}(\vx)$ is a hallmark of axion electrodynamics and  implies that the dynamical axion decays into two photons. The pseudoscalar nature
of the interaction entails a particular polarization pattern of the photons in the final state. This telltale polarization pattern suggests an observational avenue to probe the emergent  dynamical axion collective excitation  by studying the correlations of the two photons in the final state.

Motivated by its fundamental importance in topological condensed matter as well as in cosmology and particle physics, bolstered by the recent observation of dynamical axion quasiparticles in 2D topological insulators, our main objectives in this study are the following: \textbf{i:)} to obtain the two-photon quantum state arising from the decay of the dynamical axion, \textbf{ii:)} to study its polarization and entanglement properties focusing on suitable observables and correlations that characterize the telltale polarization aspects of this state, \textbf{iii:)} to explore possible observational strategies that would allow to probe the emergence of dynamical axion quasiparticles with two-photon correlations.

\vspace{1mm}

\textbf{Brief summary of results:}

To this aim we   extend the Weisskopf-Wigner formulation of spontaneous decay in atomic systems to the (effective) quantum field theory of axion (quasi) particles interacting with the electromagnetic field via a topological Chern-Simons term. This non-perturbative formulation allows us to obtain the two-photon quantum state that results from axion decay. It features a distinct pattern of momentum-polarization hyperentanglement as a result of the parity and time reversal breaking of the interaction. We introduce quantum Stokes operators to study the polarization characteristics of the two-photon state. Whereas the \emph{averages} of the degree of polarization and polarization asymmetry vanish, there are non-trivial momentum correlations of these operators which we obtain explicitly. In particular the momentum correlations of the polarization asymmetry operator are directly related to the probabilities of coincident two-photon detection which explicitly probes momentum and polarization hyperentanglement. We relate the momentum correlations of the Stokes operators to momentum and polarization resolved Hanbury-Brown Twiss second order coherences. We recognize a striking similarity with two-photon parametric down conversion and other systems  with spontaneous emission into two-photon hyperentangled states,  suggesting that experimental setups that harnessed these emission processes to test Bell inequalities  \emph{may} provide avenues to probe dynamical axion quasiparticles by measuring two-photon correlations.

The article is organized as follows: in section (\ref{sec:axions}) we introduce the effective field theory for dynamical   axions, and   the interaction picture to study time evolution to leading order in the couplings. Section (\ref{sec:ww}) presents the extension of Weisskopf-Wigner theory that describes the dynamics of spontaneous emission in multilevel atoms to study the decay of axion quasiparticles into two photons. We obtain explicitly the two photon state from axion decay and discuss momentum-polarization hyperentanglement.  In section (\ref{sec:corre}) we focus on two-photon correlations, introduce the quantum Stokes operators and obtain the averages and correlations in the two-photon state. In this section we establish a relation between the correlation of the polarization asymmetry and the coincident probability of two-photon detection. We also show that momentum correlations of the quantum Stokes operators are related to momentum and polarization resolved   Hanbury-Brown Twiss second order coherence. We recognize a striking similarity between the two-photon state from dynamical axion decay and spontaneous parametric down conversion and other condensed matter systems featuring spontaneous emission into hyperentangled two-photon states.  This similarity along with the relation between the quantum Stokes operators and  second order (HBT) coherence, suggests that experimental avenues that harness two-photon entanglement to test Bell inequalities from parametric down conversion, can be applied to topological materials  to probe the emergence of dynamical axion quasiparticles with two-photon correlations and interferometry. In section (\ref{sec:discussion}) we discuss various aspects related to the Weisskopf-Wigner formulation and   complementary aspects  of spatio-temporal correlations that merit further study.  Our conclusions and further questions are summarized in section (\ref{sec:conclusions}). Various appendices are devoted to technical aspects, including  unitarity of the Weisskopf-Wigner extension to field theory.

\section{Axionic quasiparticles.}\label{sec:axions}
The effective action for the emergent   dynamical axion field in topological insulators is given by\cite{rundong,jing,nomura}\footnote{We have absorbed the vacuum dielectric constant, vacuum permittivity  and a factor $4\pi$ into a redefinition of the gauge fields and set the speed of light $c=1$. }
\be  {S} = \int d^3x dt \Bigg\{ \frac{1}{2}\Big(  \vec{E}^2- {\vec{B}^2}  \Big)+\frac{\mathcal{J}}{2} \Bigg(\Big(\frac{\partial \delta \theta}{\partial t}\Big)^2- \Big(\vec{v}\cdot \vec{\nabla}\delta \theta\Big)^2-m^2\, \delta \theta^{\,^2} \Bigg) + \frac{\alpha}{ \pi} \,\delta \theta \,\vec{E}\cdot \vec{B}  \Bigg\}\,,\label{synac}  \ee where  the dynamical axion field $\delta \theta$ is related to the fluctuations of the Neel order parameter in the case of topological magnetic insulators, $\mathcal{J},\vec{v}$ are model and material dependent constants, $m$ is the dynamical axion mass\cite{rundong,jing,nomura}, and  $\alpha$ is the fine structure constant. Redefining the (canonically normalized) dynamical axion field
\be \phi(\vx,t) = \sqrt{\mathcal{J}}\, \delta \theta(\vx,t) \,,\label{physint}\ee and assuming rotational invariance
the effective action for the   axionic quasiparticle field and the electromagnetic field is
\be {S}  = \int d^3x dt \Bigg\{ \frac{1}{2}\Big(  \vec{E}^2- {\vec{B}^2}  \Big)+\frac{1}{2} \Bigg(\Big(\frac{\partial \phi(\vx,t)}{\partial t}\Big)^2- v^2\,\Big(  \vec{\nabla}\phi(\vx,t)\Big)^2-m^2\, \phi^{\,2}(\vx,t)  \Bigg) + g \,\phi(\vx,t) \,\vec{E}\cdot \vec{B}  \Bigg\} \,,\label{synacfin} \ee with
$  g \equiv  {\alpha}/{\pi\sqrt{\mathcal{J}}}$.  While the form of the second bracket in $S$ may differ in the realizations of  dynamical axions in topological insulators and Weyl semimetals, the coupling to electromagnetism via the last, Chern-Simons  term is a generic and distinct hallmark of the coupling of   axions to $U(1)$ gauge fields.

 The time evolution   is determined by the quantum Hamiltonian operator. Since the interaction involves a time derivative, Hamiltonian quantization is
subtle  because the Chern-Simons interaction modifies the canonical momentum of the gauge field. The proper Hamiltonian quantization procedure  along with the interaction picture representation are  discussed in detail in reference\cite{boyhybrid}.    The main result is that  to leading order in the interaction the total Hamiltonian is
\be H = H_{0EM}+H_{0\phi}+H_I \,,\label{totHam}\ee with $H_{0EM},H_{0\phi}$ are the free field Hamiltonians for electromagnetism, and axionic quasiparticle   respectively and to leading order in the couplings $g$     the interaction Hamiltonian  in  \emph{interaction picture} is given by (see ref.\cite{boyhybrid})
\be H_I(t) = - g\,\int d^3x   \phi(\vx,t)\,\vec{E}(\vx,t)\cdot\vec{B}(\vx,t) \,,\label{HIoft2}\ee
\be \vec{E}(\vx,t) = - \frac{\partial}{\partial t}\vec{A}(\vx,t) \,. \label{efield}\ee

The quantized fields in   interaction picture are expanded as
 \be \phi(\vx,t) = \frac{1}{\sqrt{V}}\,\sum_{\vk} \frac{1}{\sqrt{2E_\phi(\vk)}}\,\Big[b(\vk)\,e^{-iE_\phi(\vk)t}\,e^{i\vk\cdot \vx} + b^\dagger(\vk)\,e^{iE_\phi(\vk)t}\,e^{-i\vk\cdot \vx}  \Big]\,,\label{scalarquant2}\ee where $V$ is the quantization volume and
 \be E_\phi(k) = \sqrt{v^2\,k^2+m^2} \,,\label{energies} \ee are the single axion (quasi) particle energies. In Coulomb gauge the vector potential is given by
  \be \vec{A}(\vx,t) =  \sum_{\vk}\sum_{\lambda=1,2}  \frac{\hat{\vec{\epsilon}}_\lambda(\vk)}{\sqrt{2k V}}\, \Big[a_{\lambda}(\vk)\,e^{-ik t}\,e^{i\vk\cdot \vx} + a^\dagger_{\lambda}(\vk)\,e^{ikt}\,e^{-i\vk\cdot \vx}  \Big]\,,\label{vecpot2}\ee where the annihilation and creation operators obey the usual canonical commutation relations and the real linear polarization unit vectors $\hat{\vec{\epsilon}}_{\lambda}(\vec{k})$ are defined so that these along with the unit vector $\hat{\vec{k}}$ form a right handed triad basis with the properties
   \be \vec{\epsilon}_{1}(\vk)\times\vec{\epsilon}_{2}(\vk)= \hat{\vk}~~;~~\vec{\epsilon}_{2}(\vk)\times\hat{\vk}= \vec{\epsilon}_{1}(\vk)~~;~~\vec{\epsilon}_{1}(\vk)\times\hat{\vk}=-\vec{\epsilon}_{2}(\vk) \,, \label{polas}\ee
and chosen to satisfy the relations
\be  \vec{\epsilon}_{1}(-\vk)= - \vec{\epsilon}_{1}(\vk)~~;~~\vec{\epsilon}_{2}(-\vk)= \vec{\epsilon}_{2}(\vk)\,.\label{epsi}\ee

\section{Field theory extension of Weisskopf-Wigner theory  :}\label{sec:ww}

\subsection{General formulation:}\label{subsec:genww}

Our objective is to obtain the two-photon quantum state from the decay of the axion quasiparticle. To this aim, we begin by  extending the  formulation of spontaneous emission in multi-level atoms   based on the Weisskopf-Wigner theory of atomic linewidths  ubiquitous in quantum optics\cite{ww,zubairy,meystre} to the realm of a quantum field theory.

Let us consider the general case of a system whose Hamiltonian $H$ is given as a soluble part $H_0$ and a perturbation $H_I$: $H=H_0+H_I$. The time evolution of states in the interaction picture
of $H_0$ is given by
\be i \frac{d}{dt}|\Psi(t)\rangle_I  = H_I(t)\,|\Psi(t)\rangle_I,  \label{intpic}\ee
where the interaction Hamiltonian in the interaction picture is
\be H_I(t) = e^{iH_0\,t} H_I e^{-iH_0\,t} \label{HIoft}\,,\ee where $H_I$ is proportional to a set of couplings assumed to be small.

 The state $|\Psi(t)\rangle_I$ can be expanded as  \be |\Psi(t)\rangle_I = \sum_n C_n(t) |n\rangle \label{decom}\ee where $|n\rangle$ form a complete set of orthonormal eigenstates of $H_0$, namely $H_0\ket{n} = E_n\ket{n}$; in the many body case   these are  many-particle Fock states. From eqn.(\ref{intpic}), and the expansion (\ref{decom})  one finds the   equation of motion for the coefficients $C_n(t)$, namely

\be \dot{C}_n(t) = -i \sum_m C_m(t) \langle n|H_I(t)|m\rangle \,. \label{eofm}\ee

Although this equation is exact, it generates an infinite hierarchy of simultaneous equations when the Hilbert space of states spanned by $\{|n\rangle\}$ is infinite dimensional. However, this hierarchy can be truncated by considering the transition between states connected by the interaction Hamiltonian at a given order in $H_I$.

\begin{figure}[ht]
\includegraphics[height=3.0in,width=3.0in,keepaspectratio=true]{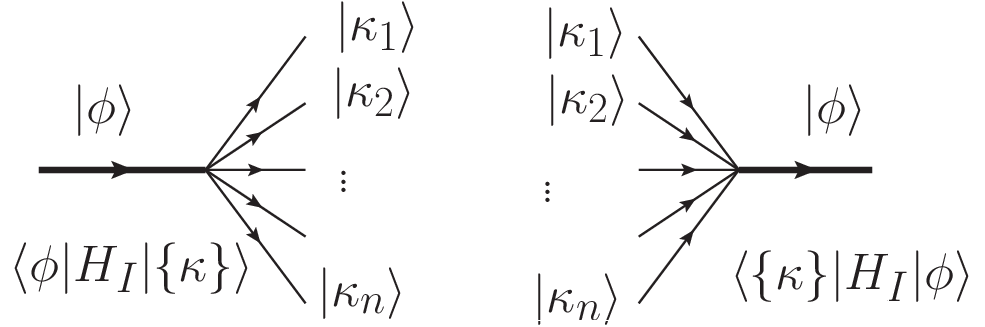}
\caption{Transitions  $\ket{\phi} \Leftrightarrow |\{\kappa_i\}\rangle \neq \ket{\phi}$,   induced by the interaction Hamiltonian  $H_I$. The set of states $|\{\kappa_i\}\rangle$ belong to a continuum.   }
\label{fig:wwtransitions}
\end{figure}

Let us consider a quantum state $\ket{\phi}$, which in the many body case may be single particle momentum eigenstates of the Fock quanta of a field, and focus on  the case when the interaction Hamiltonian does not feature a diagonal matrix element,   namely $\langle\phi|H_I|\phi\rangle =0$. Instead these states are connected to a   set of intermediate states belonging to a continuum,  $|\{\kappa\}\rangle$ by  $H_I$, namely $\ket{\phi} \Leftrightarrow |\{\kappa\}\rangle \neq \ket{\phi}$ as depicted in fig. (\ref{fig:wwtransitions}).

In the   subspace $\ket{\phi},  |\{\kappa_i\}\rangle $ the quantum state in the interaction picture is given by
\be |\Psi\rangle_I(t) =  C_{\phi}(t)\ket{\phi} \otimes \ket{\{0_{\kappa}\}}+ \sum_{\kappa_i} C_{\kappa_i}(t)  \ket{\kappa_i}    \,, \label{state}\ee
and the set of equations (\ref{eofm})   become
\bea
\dot{C}_{\phi}(t) & = & -i \sum_{\kappa_i}  \langle\phi|H_I(t)| \kappa_i \rangle \, C_{\kappa_i}(t)\, \label{eqnc1} \\
\dot{C}_{ \kappa_i}(t) & = & -i   \langle  \kappa_i |H_I(t)|\phi\rangle\, C_\phi(t) \,.\label{eqncm}\eea   where   the time dependent transition matrix elements are given by
\be \langle l|H_I(t)|m\rangle = T_{lm}\,  e^{i(E_l-E_m)t} ~~;~~ T_{lm}= \langle l|H_I(0)|m\rangle \,,\label{mtxele} \ee hermiticity of $H_I$ entails that
  \be T_{ml} = T^*_{lm} \,.\label{hermi}\ee

  The set of equations (\ref{eqnc1}-\ref{eqncm}), truncates the hierarchy of equations by neglecting the transitions between the states $|\{\kappa\}\rangle$ and  $|\{\kappa'\}\rangle \neq |\{\kappa\}\rangle,\ket{\phi}$, such transitions connect the states $\ket{\phi} \leftrightarrow |\{\kappa'\}\rangle $ at higher order in $H_I$ and are neglected up to $\mathcal{O}(H^2_I)$.  Truncating the hierarchy closes the set of equations for the amplitudes, effectively reducing the set of states to a closed subset in the full Hilbert space.

  Taking the normalized initial  quantum state $ \ket{\Psi(t=0)}$ as
\be \ket{\Psi(t=0)} =  C_{\phi}(0)\ket{\phi}  \otimes \ket{0_\kappa}  \,,\label{inistate} \ee where $\ket{0_\kappa}$ is the vacuum state for the intermediate states $\ket{\kappa}$, corresponding to initializing
\be C_{\kappa_i}(0) = 0 ~~~\forall i \,, \label{cmini}\ee  and with normalization condition
\be   C_{\phi}(0)   =1 \,. \label{unit}  \ee

The set of equations (\ref{eqncm}) with the initial condition (\ref{cmini}) can be integrated to yield
\be C_{\kappa_i}(t) = -i \int^t_0  T_{\kappa_i  \phi} \,e^{i(E_{\kappa_i}-E_\phi)t'}\, C_{ \phi}(t') \,dt' \,.\label{ckapaoft} \ee  Inserting this solution    into   equation (\ref{eqnc1}) leads to
\be
  \dot{C}_{\phi}(t)   =  -  \int^t_0  \sum_{\kappa_i}    |T_{\phi \kappa_i}|^2 \, e^{i(E_\phi-E_{\kappa_i})(t-t')}\,   {C}_{\phi}(t')   \, dt'\,, \label{dotC1} \ee

This procedure of solving for the amplitudes of  the intermediate states   plays the role of  ``tracing over'' the $\kappa$ degrees of freedom,   yielding an effective  equation   for the amplitude of  the state $\ket{\phi}$.
Since the interaction Hamiltonian $H_I$ is assumed to be proportional to weak couplings, the amplitude equations (\ref{ckapaoft},\ref{dotC1}) are exact up to second order in these couplings.

Once equation (\ref{dotC1}) is solved, the solution is inserted into equation (\ref{ckapaoft}) to find the amplitudes of the
states $\{\ket{\kappa_i}\}$, finally yielding the complete solution of the state vector (\ref{state}) in the interaction picture.  With this state, we are now in position to
obtain correlation functions of interaction picture field operators in this state to study the time dependence of correlations. This will be the strategy adopted to study
photon correlations from the decay of axionic quasiparticles.

\vspace{1mm}

\textbf{Unitarity:} The set of equations (\ref{eqnc1}-\ref{eqncm}) describe \emph{unitary time evolution} in the restricted Hilbert space of states $\Big\{\ket{\phi},|\{\kappa\}\rangle\Big\}$ which is a sub-set of the full Hilbert space of the theory that is closed under the amplitude equations   (\ref{eqnc1}-\ref{eqncm}). Unitarity can be seen as follows: using the equations (\ref{eqnc1}-\ref{eqncm}), and noticing that $\langle l|H_I(t)|m\rangle^* = \langle m|H_I(t)|l\rangle $ because $H_I(t)$ is an Hermitian operator, it follows from equations (\ref{eqnc1}-\ref{eqncm}) that
\be \frac{d}{dt} \Bigg[ | {C}_{\phi}(t)|^2 + \sum_{ \kappa_i} | {C}_{\kappa_i}(t)|^2\Bigg] =0 \,,\label{totder}\ee and the initial conditions (\ref{unit},\ref{cmini}) yield \be |{C}_{\phi}(t)|^2 + \sum_{ \kappa_i} | {C}_{\kappa_i}(t)|^2 = 1 \,.  \label{unitarity} \ee This is the statement that time evolution within the sub-Hilbert space $\Big\{\ket{\phi},|\{\kappa\}\rangle\Big\}$  is unitary, namely ${}_I\langle\Psi(t)|\Psi(t)\rangle_I =1$.

In particular if the $\ket{\phi}$ states decay, it follows that $|C_{\phi}(t=\infty)|^2=0$,  and asymptotically
\be \sum_{\kappa_i}|C_{\kappa_i}(t=\infty)|^2 = 1 \,. \label{inftyti}\ee

\textbf{Markov approximation: }

The integro-differential equation for the amplitude $C_{\phi}(t)$, equation (\ref{dotC1}), can be solved by Laplace transform: taking the Laplace transform of this equation yields
an algebraic equation for the Laplace transform of the amplitude, the inverse transform is performed in the complex plane. The solution via Laplace transform is analyzed in appendix (\ref{app:lapla}). However, we are interested in the intermediate and long-time behavior which is typically described by an exponential decay law, which can be extracted more simply by implementing a Markov-type approximation valid in the weak coupling regime.

Let us define
\be \int^{t'}_0 \sum_{\kappa_i}|T_{\phi\,\kappa_i}|^2 \, e^{i(E_\phi-E_{\kappa_i})(t-t'')}\,dt''    \equiv W[t;t']~~;~~ W[t;0]=0 \,\label{defW}\ee so that

\be \sum_{\kappa_i}|T_{\phi\,\kappa_i}|^2 \, e^{i(E_\phi-E_{\kappa_i})(t-t')}= \frac{d}{dt'}W[t;t']\,.\label{iden} \ee  Inserting this definition in (\ref{dotC1}) and integrating by parts
\be \int^t_0 \frac{d}{dt'}W[t;t']\,C_\phi(t')\,dt' = W[t;t]\,C_\phi(t)-\int^t_0  W[t;t']\,\frac{d}{dt'}C_\phi(t')\,dt'\,,\label{intparts}\ee since $W[t,t]\propto |T_{\phi\,\kappa_i}|^2 \propto H^2_I$ and from the amplitude equation (\ref{dotC1}) it follows that $\dot{C}_\phi \propto H^2_I$,    the second term (integral) on the right hand side in (\ref{intparts}) is of $\mathcal{O}(H^4_I)$ and will be neglected to leading order in the interaction, namely $H^2_I$.

 Hence, up to $\mathcal{O}(H^2_I)$, the evolution equation for the amplitude (\ref{dotC1}) becomes
\be \dot{C}_\phi(t)  =  -  W[t;t]\,C_\phi(t)~~;~~C_\phi(0)=1\,.\label{dotc1fin} \ee Following the main Markov approximation in the Weisskopf-Wigner method\cite{ww,zubairy,meystre} we  invoke the
 long time limit
 \be \int^{t}_0    e^{i(E_\phi-E_{\kappa_i})(t-t' )}\,dt'~{}_{\overrightarrow{t\rightarrow \infty }}~ i\,\Bigg[\mathcal{P}\,\Big(\frac{1}{E_\phi-E_{\kappa_i}} \Big) -i \pi \delta(E_\phi-E_{\kappa_i}) \Bigg]\,,\label{pp} \ee and let us introduce     the spectral density
\be \rho (k_0) =  \,\sum_{\kappa_i} |T_{\phi\,\kappa_i}|^2 \,\delta(k_0-E_{\kappa_i}) \,, \label{rhosd}\ee    yielding
 \be \,W[t;t]~{}_{\overrightarrow{t\rightarrow \infty }}~i\,\Sigma(E_\phi) \,. \label{Ws} \ee The self-energy
 \be \Sigma(E_\phi)  = \int^{\infty}_{-\infty} \frac{\rho(k_0)}{E_\phi-k_0+ i\varepsilon} \,dk_0   ~~;~~ \varepsilon \rightarrow 0^+ \label{selfenergy}\ee has a simple and intuitive interpretation as a second order Feynman diagram wherein the lines representing the  intermediate states $\{\ket{\kappa_i}\}$ in fig. (\ref{fig:wwtransitions}) are joined into ``propagators'' yielding a  multi- loop diagram,   representing the self-energy up to second order in $H_I$.

Separating the self-energy into the real and imaginary parts
\be  \Sigma (E_\phi)  = \Delta -i \frac{\Gamma  }{2} ~~;~~ \Delta  \equiv \int^{\infty}_{-\infty}\mathcal{P} \Bigg[  \frac{\rho (k_0)}{E_\phi-k_0} \Bigg]\,dk_0 ~~;~~ \Gamma  = 2\pi\,\rho(E_\phi) \,,\label{RIsig}\ee where $\mathcal{P}$ stands for the principal part.   $\Delta$ is the energy renormalization (Lamb shift) and $\Gamma$ the decay rate or line-width as per Fermi's Golden rule. Therefore the solution of the amplitude equation (\ref{dotc1fin})   is
\be C_\phi(t) = e^{-i\Delta t}\,e^{-\frac{\Gamma}{2} t} \,,\label{cfisol} \ee and inserting this solution into equation (\ref{ckapaoft}) yields the amplitudes
\be C_{\kappa_i}(t) = T_{\kappa_i  \phi}\,\Bigg[\frac{1-e^{-i(E^r_{\phi}-E_{\kappa_i}-i\frac{\Gamma}{2})t}}{E_{\kappa_i}-E^r_{\phi}+i\frac{\Gamma}{2}}\Bigg]\,, \label{ckapaifin}\ee where
\be E^r_{\phi} = E_{\phi}+\Delta \,,\label{efiren}\ee is the renormalized energy. We can now confirm the validity of the unitarity relation (\ref{inftyti}) up to second order in
the interaction
\be \sum_{\kappa_i} |C_{\kappa_i}(\infty)|^2 = \frac{2\pi}{\Gamma} \,\sum_{\kappa_i} |T_{\kappa_i  \phi} |^2\, \frac{1}{\pi} \frac{\Gamma/2}{(E^r_{\phi}-E_{\kappa_i})^2+(\Gamma/2)^2} ~~{}_{\overrightarrow{\Gamma \rightarrow 0}} ~~  \frac{2\pi}{\Gamma}\,\sum_{\kappa_i} |T_{\kappa_i  \phi} |^2\,\delta(E_{\phi}-E_{\kappa_i}) =1 \,,\label{uniconf} \ee where, consistently with the weak coupling expansion,  we used the narrow width limit $\Gamma\rightarrow 0$ to replace the Lorentzian by the delta function, the difference being of $\mathcal{O}(H^2_I)$ which can be neglected since it multiplies $|T|^2\simeq \mathcal{O}(H^2_I)$,  along with the relations (\ref{rhosd},\ref{RIsig})  and replaced $E^r_\phi\rightarrow E_{\phi}$,  to leading
order  ($\mathcal{O}(H^2_I)$) in the coupling.  Keeping the full time dependence of $W[t,t]$ (beyond Markov) is discussed in appendix   (\ref{subsec:nonmarkov}).

\subsection{Axionic quasiparticles coupled to photons:}\label{subsec:axiphot}
We now apply the   Weisskopf-Wigner formulation described above to the case of axionic quasiparticles coupled to electromagnetism via the Chern-Simons term,
described by the total Hamiltonian (\ref{totHam}) with the interaction term given by (\ref{HIoft2}).  With the field expansions (\ref{scalarquant2},\ref{vecpot2}) the interaction Hamiltonian in the interaction picture becomes
\bea H_I(t) & = &   {g} \sum_{\vp} \sum_{\stackrel{\vp_1,\lambda_1}{\vp_2,\lambda_2}} \sqrt{\frac{p_1p_2}{8VE_{\phi}(p)}}\,\Big[\hat{\vec{\epsilon}}_{\lambda_1}(\vp_1)\cdot\big(\hat{\vp}_2\times \hat{\vec{\epsilon}}_{\lambda_2}(\vp_2) \big)\Big] \Bigg[ b(\vp) \, a^\dagger_{\lambda_1}(\vp_1) a^\dagger_{\lambda_2}(\vp_2) \,e^{-i(E_\phi(p)-p_1-p_2)t} + \emph{h.c.} \nonumber \\ & + &  \cdots \Bigg]\delta_{\vp_1+\vp_2,\vp}\,.\label{HIcs} \eea In the following analysis we will neglect the terms $(\cdots)$ in eqn. (\ref{HIcs}), these do not yield resonant, energy conserving contributions to leading order. Keeping solely the terms displayed
in the first line in (\ref{HIcs}) is tantamount to the rotating wave approximation (RWA), ubiquitous  in quantum optics\cite{meystre,zubairy}. The contributions of non-(RWA) terms are disconnected vacuum diagrams, which are neglected in the following discussion whose focus is on the two-photon final state,  but will be discussed in more detail in section (\ref{sec:discussion}).

\begin{figure}[ht]
\includegraphics[height=3.0in,width=3.0in,keepaspectratio=true]{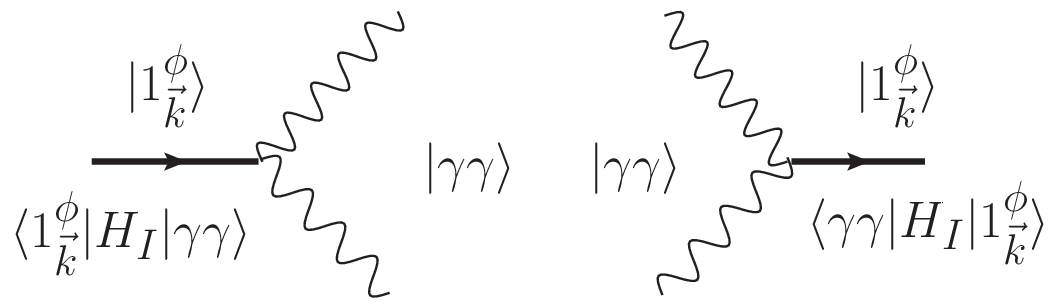}
\caption{Transitions  $\ket{1^{\phi}_{\vk}} \Leftrightarrow \ket{\gamma\gamma} \equiv \ket{1_{\vk_1,\lambda_1};1_{\vk_2,\lambda_2}} $, with $\vk_2=\vk-\vk_1$  induced by the interaction Hamiltonian  $H_I$ (\ref{HIcs}).   }
\label{fig:transitions}
\end{figure}

Let us consider the decay of a single   axionic (quasi) particle, identifying the state $\ket{\phi}$ in   (\ref{state}) with the single particle  Fock state $\ket{1^{\phi}_{\vk}}$. It is clear from the interaction Hamiltonian that the intermediate set of states, namely  $\ket{\kappa_i}$ in (\ref{state}),  coupled to this single particle state are two photons states of the form $\ket{1_{\vk_1,\lambda_1};1_{\vk_2,\lambda_2}}$, depicted in figure (\ref{fig:transitions}).

 The transition matrix elements (\ref{mtxele})  are given by
\be T_{\phi \gamma\gamma} \equiv \langle 1^{\phi}_{\vk}|H_I(0)|1_{\vk_1,\lambda_1};1_{\vk_2,\lambda_2}\rangle =  g  \, \sqrt{\frac{k_1k_2}{8VE_{\phi}(k)}}\,\Big[\hat{\vec{\epsilon}}_{\lambda_1}(\vk_1)\cdot\big(\hat{\vk}_2\times \hat{\vec{\epsilon}}_{\lambda_2}(\vk_2) \big)+\hat{\vec{\epsilon}}_{\lambda_1}(\vk_2)\cdot\big(\hat{\vk}_1\times \hat{\vec{\epsilon}}_{\lambda_1}(\vk_1) \big) \Big]\,  \delta_{\vk_1+\vk_2,\vk} \,.\label{Tfig}\ee

We can now implement the general results obtained in the previous section to obtain the time evolved state in the interaction picture (\ref{state}), by identifying  $\ket{\phi}\rightarrow \ket{1^{\phi}_{\vk}}~; ~\ket{\kappa_i} \rightarrow \ket{1_{\vk_1,\lambda_1};1_{\vk_2,\lambda_2}}$. However, in this case there is a subtle but important aspect, the final two photon state features indistinguishability between the two photons, therefore the time evolved state in the interaction picture is
\be |\Psi(t)\rangle_I  =   C_\phi(t) \ket{1^{\phi}_{\vk}} \otimes \ket{\{0_{\gamma}\}}+ \ket{\psi_{\gamma\gamma}(t)}\otimes\ket{0_\phi}    \,, \label{statefin}\ee where the two-photon state
\be \ket{\psi_{\gamma\gamma}(t)} = \frac{1}{2\,!} \sum_{\vk_1,\lambda_1}\sum_{\vk_2,\lambda_2} C_{\gamma\gamma}(\vk_1,\lambda_1;\vk_2,\lambda_2;t)  \ket{1_{\vk_1,\lambda_1};1_{\vk_2,\lambda_2}}\,.\label{2phots}\ee The amplitude $C_{\gamma\gamma}(\vk_1,\lambda_1;\vk_2,\lambda_2;t)$ is symmetric under the exchange
$\vk_1,\lambda_1 \Leftrightarrow \vk_2,\lambda_2$ and the factor $1/2!$ compensates for the double counting as a consequence of the indistinguishability of the two photon state
and the concomitant symmetry of the coefficient. Using that
 \be \langle 1_{\vp_1,\alpha_1};1_{\vp_2,\alpha_2}|1_{\vk_1,\lambda_1};1_{\vk_2,\lambda_2} \rangle = \delta_{\vp_1,\vk_1}\delta_{\alpha_1,\lambda_1}\delta_{\vp_2,\vk_2}\delta_{\alpha_2,\lambda_2}+
 \delta_{\vp_2,\vk_1}\delta_{\alpha_2,\lambda_1}\delta_{\vp_1,\vk_2}\delta_{\alpha_1,\lambda_2}\,,\label{comi}\ee
the amplitude equations (\ref{eqnc1},\ref{eqncm}) become

 \bea
& & \dot{C}_{\phi}(t)  =  -\frac{i}{2} \sum_{\vk_1,\lambda_1}\sum_{\vk_2,\lambda_2}  \langle\phi|H_I(t)| 1_{\vk_1,\lambda_1};1_{\vk_2,\lambda_2} \rangle \, C_{\gamma\gamma}(\vk_1,\lambda_1;\vk_2,\lambda_2;t) ~;~~ {C}_{\phi}(0)=1\, \label{eqncfi} \\
& & \dot{C}_{\gamma\gamma}(\vk_1,\lambda_1;\vk_2,\lambda_2;t)  =  -i   \langle   1_{\vk_1,\lambda_1};1_{\vk_2,\lambda_2} |H_I(t)|\phi\rangle\, C_\phi(t) ~;~~ {C}_{\gamma\gamma}(\vk_1,\lambda_1;\vk_2,\lambda_2;0)=0 \,.\label{eqnckas}\eea
Implementing the steps leading up to equations (\ref{ckapaoft},\ref{dotC1}) we now find
\be C_{\gamma\gamma}(\vk_1,\lambda_1;\vk_2,\lambda_2;t) = -i \int^t_0  T_{\gamma\gamma  \phi} \,e^{i(E_{\gamma\gamma}-E_\phi)t'}\, C_{ \phi}(t') \,dt' \,.\label{cgaga} \ee  and
\be
  \dot{C}_{\phi}(t)   =  -  \int^t_0  \sum_{\vk_1,\lambda_1}\sum_{\vk_2,\lambda_2}   \frac{1}{2!}\, |T_{\phi \gamma \gamma}|^2 \, e^{i(E_\phi-E_{\gamma\gamma})(t-t')}\,   {C}_{\phi}(t')   \, dt'\,, \label{cifit} \ee
 where $T_{\phi \gamma\gamma} = T^*_{\gamma \gamma \phi}$ is given by the result (\ref{Tfig}) and
 \be E_{\gamma\gamma} = k_1+k_2 \,. \label{egaga} \ee

\begin{figure}[ht]
\includegraphics[height=2.5in,width=2.5in,keepaspectratio=true]{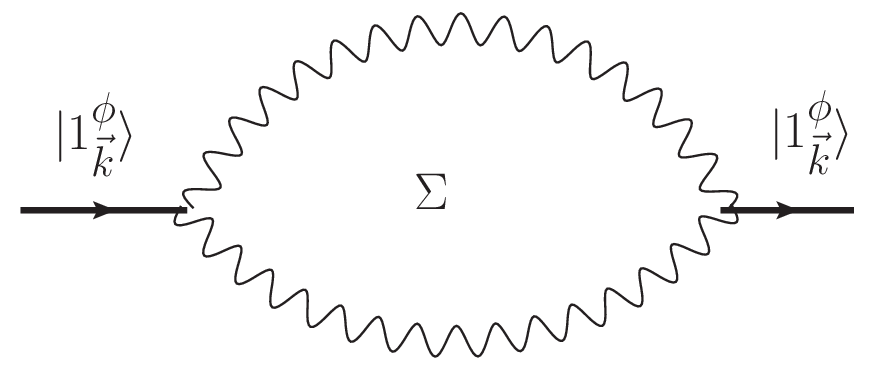}
\caption{Second order self-energy diagram.   }
\label{fig:onelup}
\end{figure}

  Now, including the symmetry factor accounting for the indistinguishability of the two photon state,  the spectral density  (\ref{rhosd}) is given by
\be \rho(k_0) = \sum_{\vk_1,\lambda_1}\sum_{\vk_2,\lambda_2} \frac{1}{2!}\, |T_{\phi \gamma\gamma}|^2\,\delta(k_0-k_1-k_2) \,\label{rhoax}\ee from which the self-energy $\Sigma$ is obtained
via the dispersive  (Kramers-Kronig) representation  (\ref{selfenergy}). To leading order  in the axion-photon coupling $\Sigma$ is represented by  the one-loop Feynman diagram shown in figure (\ref{fig:onelup}).

Passing to the $V\rightarrow \infty$ limit and summing over polarizations $\rho(k_0)$ becomes (see appendix (\ref{app:spec}))

\be \rho(k_0) = \frac{g^2}{8 E_\phi(k)}\,\int\frac{d^3k_1}{(2\pi)^3}\frac{1}{k_1|\vec{k}-\vec{k}_1|}\,\Big[k_1|\vec{k}-\vec{k}_1|-\vk_1\cdot(\vk-\vk_1) \Big]^2 \,\delta(k_0-k_1-|\vec{k}-\vec{k}_1|) \,,\label{rhoaxfin}\ee

The details of its calculation are provided in appendix (\ref{app:spec}), with the result
\be \rho(k_0) = \frac{g^2}{64\pi^2 E_{\phi}(k)}\Big(k^2_0-k^2 \Big)^2\,\Theta(k_0-k)\,, \label{rhoabfinal} \ee yielding
for the real and imaginary parts of the self-energy  $\Sigma$ (see eqn. (\ref{RIsig}))
\bea \Delta(k)  & = & \frac{g^2}{64\pi^2 E_\phi(k)}\int^{\Lambda}_k \mathcal{P} \Big[ \frac{(k^2_0-k^2)^2}{E_\phi(k)-k_0} \Big]\,dk_0  \,,\label{Deltafi}\\  \Gamma & =  & 2\pi\,\rho(E_{\phi})=  \frac{g^2\,(E^2_\phi(k)-k^2)^2}{32\pi E_\phi(k)}\,\,\Theta(E_{\phi}(k)-k)\,,   \label{Gamafi}\eea where we have introduced an upper cutoff $\Lambda$ in the integrals for the real part  $\Delta(k) $ (Lamb-shift).

We note the Cerenkov-like condition
\be (1-v^2) < \frac{m^2}{k^2} \,,\label{cerecon}\ee for the axionic quasiparticle to decay into two photons.

Implementing the Weisskopf-Wigner formulation described above, we now find (in the Markov approximation)
\be C_\phi(t) =  e^{-i\Delta(k) t}\,e^{-\frac{\Gamma}{2} t} \,,\label{cifitsol}\ee and the amplitudes for the two-photon state (\ref{2phots}) are given by
\be C_{\gamma\gamma}(\vk_1,\lambda_1;\vk_2,\lambda_2;t)  =   g\, \sqrt{\frac{k_1k_2}{8 V E_{\phi}(k)}}\,\mathcal{P}(\vk_1,\lambda_1;\vk_2,\lambda_2) \,\Bigg[\frac{1-e^{-i(E^r_{\phi}-E_{\gamma\gamma}-i\frac{\Gamma}{2})t}}{E_{\gamma\gamma}-E^r_{\phi}+i\frac{\Gamma}{2}}\Bigg]\,\delta_{\vk_1+\vk_2,\vk}\,,\label{cgg} \ee where
\be  E^r_{\phi}(k) = E_\phi(k)+\Delta(k) \,, \label{ees}\ee is the renormalized axion quasiparticle energy including the Lamb-shift $\Delta(k)$, and
\be \mathcal{P}(\vk_1,\lambda_1;\vk_2,\lambda_2) \equiv \Big[\hat{\vec{\epsilon}}_{\lambda_1}(\vk_1)\cdot\big(\hat{\vk}_2\times \hat{\vec{\epsilon}}_{\lambda_2}(\vk_2) \big)+\hat{\vec{\epsilon}}_{\lambda_2}(\vk_2)\cdot\big(\hat{\vk}_1\times \hat{\vec{\epsilon}}_{\lambda_1}(\vk_1) \big) \Big]~~;~~ \vk_2=\vk-\vk_1\,.\label{polac}\ee  As anticipated  the amplitude (\ref{cgg}) is symmetric under the exchange $\vk_1,\lambda_1 \Leftrightarrow \vk_2,\lambda_2$,

Now the amplitude equations (\ref{eqncfi},\ref{eqnckas}) including the statistical factor $1/2!$ yield the unitarity condition
\be \frac{d}{dt}\Big[ |C_\phi(t)|^2 +  \frac{1}{2!} \sum_{\vk_1,\lambda_1}\sum_{\vk_2,\lambda_2} |C_{\gamma\gamma}(\vk_1,\lambda_1;\vk_2,\lambda_2;t)|^2\Big]=0 \,,\label{unics}   \ee which, by dint of the initial conditions on the amplitudes implies

\be   |C_\phi(t)|^2 +  \frac{1}{2!} \sum_{\vk_1,\lambda_1}\sum_{\vk_2,\lambda_2} |C_{\gamma\gamma}(\vk_1,\lambda_1;\vk_2,\lambda_2;t)|^2 =1 \,.\label{unitycs}   \ee This result is a manifestation of  the unitarity relation
\be {}_I{\langle \Psi(t)|\Psi(t)\rangle_I} =1 \,,\label{unipsi}\ee where $|\Psi(t)\rangle_I$ is the quantum state (\ref{statefin}), which is confirmed  upon using the identity (\ref{comi}) and the symmetry of the amplitude of the two-photon state. Unitarity at all time is proven explicitly in appendix (\ref{app:alltime}).

The two photon state
\be \ket{\psi_{\gamma\gamma}(t)} =   \frac{1}{2\,!} \sum_{\vk_1,\lambda_1}\sum_{\vk_2,\lambda_2} C_{\gamma\gamma}(\vk_1,\lambda_1;\vk_2,\lambda_2;t)  \ket{1_{\vk_1,\lambda_1};1_{\vk_2,\lambda_2}}\,\label{2fots}\ee
features \emph{hyperentanglement } in momentum and polarization. It is convenient to separate the kinematic and polarization components of the two-photon amplitude by writing
  \be C_{\gamma\gamma}(\vk_1,\lambda_1;\vk_2,\lambda_2;t)  \equiv \mathcal{K}( k_1;k_2;t) \times \mathcal{P}(\vk_1,\lambda_1;\vk_2,\lambda_2)\,,\label{kinpol}\ee where
  \be \mathcal{K}(k_1;k_2;t)=   g\, \sqrt{\frac{k_1k_2}{8 V E_{\phi}(k)}} \,\Bigg[\frac{1-e^{-i(E^r_{\phi}-E_{\gamma\gamma}-i\frac{\Gamma}{2})t}}{E_{\gamma\gamma}-E^r_{\phi}+i\frac{\Gamma}{2}}\Bigg]~~;~~E_{\gamma\gamma}=k_1+k_2~~;~~k_2=|\vk-\vk_1|\,,\label{kinem}\ee is completely determined by the kinematics of the decay process,
    $\mathcal{P}(\vk_1,\lambda_1;\vk_2,\lambda_2)$ is given by eqn. (\ref{polac}) and contains all the polarization information of this state.

 Well after the single axion state decayed, the asymptotic state  is given by
 \be  \ket{\Psi(\infty)}_I=\ket{\psi_{\gamma\gamma}(\infty)} =   \frac{1}{2\,!} \sum_{\substack{\vk_1,\lambda_1 \\ \vk_2,\lambda_2}} C_{\gamma\gamma}(\vk_1,\lambda_1;\vk_2,\lambda_2;\infty)  \ket{1_{\vk_1,\lambda_1};1_{\vk_2,\lambda_2}}\,.\label{2fotasy}\ee

This is one of the important results of this study and a direct bonus of the extension of the Weisskopf-Wigner theory to the quantum field theory of axion-quasiparticles interacting with photons. The fulfillment of unitarity is an important ingredient because we will obtain correlation functions of operators in the two-photon final state.

 \section{Two-photon correlations.}\label{sec:corre}

 The probability of coincident observation of  two correlated photons with momenta and polarizations $\vk_1,\lambda_1;\vk_2,\lambda_2$ respectively ($\vk_2=\vk-\vk_1$)  at time $t$  is given by
\be \mathbb{P}_{\gamma\gamma}(\vk_1,\lambda_1;\vk_2,\lambda_2;t) = \Big|\langle 1_{\vk_1,\lambda_1};1_{\vk_2,\lambda_2}|\psi_{\gamma\gamma}(t)\rangle\Big|^2  = |C_{\gamma \gamma}(\vk_1,\lambda_1;\vk_2,\lambda_2;t )|^2\,.\label{2fotprob1}\ee In particular at long time after the axionic quasiparticle has decayed it is given by

\be \mathbb{P}_{\gamma\gamma}(\vk_1,\lambda_1;\vk_2,\lambda_2)= 2\pi \frac{g^2 k_1 k_2}{8VE_{\phi}(k)\,\Gamma}\,\Big[\mathcal{P}(\vk_1,\lambda_1;\vk_2,\lambda_2)\Big]^2\, \frac{1}{\pi}\,\Bigg[ \frac{\Gamma/2}{\big(k_1+k_2-E^r_{\phi}\big)^2+\frac{\Gamma^2}{4}}  \Bigg]~~;~~ \vk_2=\vk_1-\vk\,.\label{2fotprob2}\ee

 The axion mass and its dispersion relation enter   in the denominator of the Lorentzian factor in (\ref{2fotprob2}) via $E_{\phi}$ and in the width of the Lorentzian
 distribution, $\Gamma$ via the decay width (inverse lifetime) given by eqn. (\ref{Gamafi}). The Lorentzian profile of the probability is a consequence of the time-energy uncertainty of the decaying state, a common feature of spontaneous emission, but the true telltale of two-photon correlations is completely determined by the polarization function
 $\mathcal{P}(\vk_1,\lambda_1;\vk_2,\lambda_2)$ given by equation (\ref{polac}, which characterizes momentum-polarization hyperentanglement. The distinct feature of polarization described by $\mathcal{P}(\vk_1,\lambda_1;\vk_2,\lambda_2)$  is a direct consequence of the topological and parity and time-reversal breaking nature of the Chern-Simons interaction and distinguishes topological materials from trivial ones.

 For weak coupling, this probability distribution is strongly peaked at $E_{\gamma\gamma}=k_1+k_2 = E^r_{\phi}(k)$, with $\vk_1+\vk_2=\vk$,  which are recognized as the conditions for
energy and momentum conservation in the decay. The broadening of the two-photon spectral distribution by the width $\Gamma$ (instead of a delta function) is a consequence of the energy-time uncertainty associated with the lifetime of the axion quasiparticle.

To see more clearly the type of momentum-polarization correlation, let us consider that the axionic quasiparticle features $\vk=0$, namely $\vk_2=-\vk_1$, and that one observer (Bob)
detects a photon with momentum and polarization $\vk_1,\lambda_1$ and determines a set of axis perpendicular to $\vk_1$ identifying the polarization states $\hat{\vec{\epsilon}}_1(\vk_1), \hat{\vec{\epsilon}}_2(\vk_1)$ with these axis, and another observer (Alice) measures the photon emitted along the opposite direction, with $-\vk_1$ and measures the polarization using the same axis as Bob's setup. The relations (\ref{polas},\ref{epsi}) entail the following values for $\mathcal{P}(\vk_1,\lambda_1;\vk_2,\lambda_2)$
\be \mathcal{P}(\vk_1,1;-\vk_1,1) = \mathcal{P}(\vk_1, 2;-\vk_1,2)=0~;~\mathcal{P}(\vk_1,1;-\vk_1,2)=\mathcal{P}(\vk_1,2;-\vk_1,1)=2\,.\label{polacorre}\ee

This result is a distinct hallmark of the Chern-Simons interaction: for zero momentum transfer, photons are emitted back-to-back with perpendicular polarizations, a clear
observational pattern which is  a direct consequence of the breaking of parity and time reversal invariance by the interaction\footnote{Incidentally, this is the same as the photon polarization pattern emerging from the decay at rest of a neutral $\Pi^0$  meson (pion), which just like the axion is a pseudoscalar (quasi) particle. It couples to electromagnetism similarly to the axion as a consequence of the Adler-Bell-Jackiw anomaly.}.  Each pair of back-to-back photons is featured in the two photon quantum state with a contribution
\be \mathcal{K}(k_1;k_1;t) \Big[ \ket{1_{\vk_1,1}; 1_{-\vk_1,2}}+\ket{1_{\vk_1,2}; 1_{-\vk_1,1}}\Big] \,,\label{bell}\ee which is a typical Bell state with momentum-polarization hyperentanglement.

To obtain $\mathcal{P}(\vk_1,\lambda_1;\vk_2,\lambda_2)$ for arbitrary $\vk$, with $\vk_2=\vk-\vk_1$, it is convenient to express the unit vectors $\hat{\vk}_2, \hat{\vec{\epsilon}}_1(\vk_2),\hat{\vec{\epsilon}}_2(\vk_2)$    in the basis of the orthonormal right handed triad ( $\hat{\vec{\epsilon}}_1(\vk_1),\hat{\vec{\epsilon}}_2(\vk_1),\hat{\vk}_1)$ (identified with the cartesian coordinate basis $(\hat{\vec{x}},\hat{\vec{y}},\hat{\vec{z}})$), namely

\bea \hat{\vk}_2 & = & \hat{\vec{\epsilon}}_1(\vk_1) \sin(\theta)\cos(\varphi)+ \hat{\vec{\epsilon}}_2(\vk_1)\sin(\theta)\sin(\varphi)+\hat{\vk}_1\cos(\theta) \nonumber \\
\hat{\vec{\epsilon}}_1(\vk_2) & = & \hat{\vec{\epsilon}}_1(\vk_1) \cos(\theta)\cos(\varphi)+ \hat{\vec{\epsilon}}_2(\vk_1)\cos(\theta)\sin(\varphi)-\hat{\vk}_1\sin(\theta) \nonumber \\
\hat{\vec{\epsilon}}_2(\vk_2) & = & -\hat{\vec{\epsilon}}_1(\vk_1)  \sin(\varphi)+ \hat{\vec{\epsilon}}_2(\vk_1) \cos(\varphi)\,.\label{k2vecs}\eea This basis fulfills all the relations (\ref{polas}), and the particular case $\vk=\vec{0}$, corresponds to $\theta =\pi;\varphi=0$,   so that the (right-handed) triad corresponding to $\vk_2=-\vk_1$ agrees with the choice (\ref{epsi})\footnote{Of course the choice (\ref{epsi})   can be generalized by a  rotation around the $\vk$ axis by an arbitary angle $\varphi$. }.

It is straightforward to show that the relations (\ref{polas}) for both set of triads ($ \hat{\vk}_2,\hat{\vec{\epsilon}}_1(\vk_2),
\hat{\vec{\epsilon}}_2(\vk_2)$; $ \hat{\vk}_1,\hat{\vec{\epsilon}}_1(\vk_1),
\hat{\vec{\epsilon}}_2(\vk_1) $ ) imply that
\be  \mathcal{P}(\vk_1,1;\vk_2,1) +\mathcal{P}(\vk_1, 2;\vk_2,2) =0 \,.\label{sumpolis}\ee

 We find for the general case
\bea   \mathcal{P}(\vk_1,1;\vk_2,1)  & = &  -\mathcal{P}(\vk_1, 2;\vk_2,2) = -\sin(\varphi)\, (1-\cos(\theta)) \nonumber \\   \mathcal{P}(\vk_1,1;\vk_2,2) & = & \mathcal{P}(\vk_1,2;\vk_2,1)=\cos(\varphi)\,(1-\cos(\theta))\,.\label{polacorregen}\eea In particular, it follows from this general result that
\be \sum_{\lambda_2=1}^2  \Big[\mathcal{P}(\vk_1,\lambda_1;\vk_2,\lambda_2) \Big]^2 = \sum_{\lambda_1=1}^2  \Big[\mathcal{P}(\vk_1,\lambda_1;\vk_2,\lambda_2) \Big]^2 = \Big(1-\hat{\vk}_1\cdot\hat{\vk}_2 \Big)^2\, ,  \label{unasuma}\ee and as a corollary,
\be \sum_{\lambda_1=1}^2\sum_{\lambda_2=1}^2 \Big[\mathcal{P}(\vk_1,\lambda_1;\vk_2,\lambda_2) \Big]^2 = 2\,\Big(1-\hat{\vk}_1\cdot\hat{\vk}_2 \Big)^2\,,\label{dossumasP} \ee confirming the analysis in appendix (\ref{app:spec}).

 Taken together, the results (\ref{sumpolis},\ref{polacorregen}-\ref{dossumasP}) are of observational relevance since these are independent of the choice of basis and are a distinct hallmark of the polarization pattern emerging
from the parity and time reversal non-invariant interaction of axion quasiparticles with the electromagnetic field via the Chern-Simons term.

\subsection{Stokes operators}\label{subsec:stokes}
The  Stokes parameters provide a description of the polarization properties of \emph{classical} light, conceptualized in terms
of a vector on a Poincare sphere\cite{born,mandl}. For \emph{quantum} light, Stokes operators have been introduced that provide operator representations of the   polarization degrees
of freedom in quantum optics\cite{jauch,agarwal,stokes1,qupola1}.  The Stokes operators    are generically introduced for monochromatic beams of photons of momentum $\vp$ corresponding to a single momentum mode   of the
photon quantum field (\ref{vecpot2}) with polarizations $\hat{\vec{\epsilon}}_{s}(\vec{p})~;~s=1,2$.  In terms of creation and annihilation operators of photons with momentum $\vec{p}$ and polarization $s=1,2$ the quantum Stokes operators are\footnote{In the literature   the factor $1/2$ is often omitted.}
\bea \mathcal{S}_0(\vp) & = &  \frac{1}{2}\Big(a^\dagger_1(\vp)a_1(\vp)+a^\dagger_2(\vp)a_2(\vp) \Big)\,,\label{S0}  \\
\mathcal{S}_1(\vp) & = &  \frac{1}{2}\Big(a^\dagger_1(\vp)a_1(\vp)-a^\dagger_2(\vp)a_2(\vp) \Big)\,,\label{S1} \\
\mathcal{S}_2(\vp) & = &  \frac{1}{2}\Big(a^\dagger_1(\vp)a_2(\vp)+a^\dagger_2(\vp)a_1(\vp) \Big)\,,\label{S2} \\
\mathcal{S}_3(\vp) & = &  -\frac{i}{2}\Big(a^\dagger_1(\vp)a_2(\vp)-a^\dagger_2(\vp)a_1(\vp) \Big)\,,\label{S3} \eea the spatial components obey the $SU(2)$ algebra
\be [\mathcal{S}_i(p),\mathcal{S}_j(p)] = i \epsilon_{ijk}\mathcal{S}_k(p)\,,\label{su2}\ee
\be [\mathcal{S}_0,\mathcal{S}_j] = 0~~;~ j=1,2,3 \,,\label{iden2}\ee
along with the identity
\be \mathcal{S}_0(\vp)(\mathcal{S}_0(\vp)+1) = \mathcal{S}^2_1(\vp)+ \mathcal{S}^2_2(\vp)+ \mathcal{S}^2_3(\vp)\,.\label{idenSs}\ee

The operator $S_0(\vp)$ is associated with the average photon number per polarization degree of freedom, and $S_1(\vp)$ with the \emph{polarization asymmetry}.

The \emph{classical} Stokes parameters are the expectation values of these Stokes operators in coherent states of the electromagnetic field\cite{agarwal}. Introducing the matrix
\be \mathcal{J} =  \left(
                            \begin{array}{cc}
                              \langle a^\dagger_1(\vp)a_1(\vp) \rangle & \langle a^\dagger_1(\vp)a_2(\vp) \rangle \\
                              \langle a^\dagger_2(\vp)a_1(\vp) \rangle & \langle a^\dagger_2(\vp)a_2(\vp) \rangle \\
                            \end{array}
                          \right)
 \,, \label{matJ}\ee where the expectation values $\langle (\cdots) \rangle$ are in the corresponding quantum state,  the degree of polarization is defined\cite{agarwal,qupola1} as
 \be \mathcal{D}(\vp) = \sqrt{1- \frac{4\,det{\mathcal{J}}}{\big(Tr \mathcal{J} \big)^2}}\,,\label{depol}\ee   it is  the quantum counterpart of the classical definition of degree of polarization in terms of the classical Stokes parameters\cite{born,mandl}. In \emph{average}, a (monochromatic) beam is fully polarized if $\mathcal{D} =1$ and completely un-polarized if $\mathcal{D}=0$. The diagonal matrix elements of $\mathcal{J}$ give the occupation number of photons of momentum $\vp$ and polarizations $s=1,2$ respectively.
Whereas for classical light the Stokes parameters provide a complete characterization of   polarization,
 at the quantum level, expectation values are \emph{not} sufficient to characterize the polarization because higher order correlation functions of the quantum Stokes operators also yield information on the polarization state, in particular fluctuations in polarization\cite{agarwal,qupola1,aga}. Important polarization correlations even with vanishing degree of polarization may result as a consequence of two-photon entanglement\cite{korolkova}.

 In the case under consideration, the quantum state is the two-photon state $\ket{\psi_{\gamma \gamma}(t)}$ given by eqn. (\ref{2phots}) with the amplitudes given by eqn. (\ref{cgg}). This state is certainly not monochromatic, and as discussed above, it features two-photon correlations as a consequence of momentum and polarization entanglement. Therefore we expect non-trivial correlations between Stokes operators associated with photons with different momenta and polarizations. The matrix of expectation values of the Stokes operators in the two-photon quantum state $\ket{\psi_{\gamma \gamma}(t)}$, namely $\mathcal{J}$ given by eqn. (\ref{matJ})  requires the generic matrix element
 \bea \mathcal{J}_{\alpha'\alpha}(\vp;t) \equiv \langle \psi_{\gamma \gamma}(t)| a^\dagger_{{\vp}\,\alpha'}\,a_{\vp\, \alpha}   | \psi_{\gamma \gamma}(t)\rangle & = & \frac{1}{4} \sum_{\substack{\vk'_1,\lambda'_1 \\ \vk'_2,\lambda'_2}} C^*_{\gamma\gamma}(\vk^{\,'}_1,\lambda'_1;\vk^{\,'}_2,\lambda'_2;t)  \sum_{\substack{\vk_1,\lambda_1 \\ \vk_2,\lambda_2}} C_{\gamma\gamma}(\vk_1,\lambda_1;\vk_2,\lambda_2;t)   \nonumber \\
  & \times &  \langle 1_{\vk_1',\lambda_1'};1_{\vk_2',\lambda_2'}| a^\dagger_{{\vp}\,\alpha'}\,a_{\vp \,\alpha}    |  1_{\vk_1,\lambda_1};1_{\vk_2,\lambda_2}\rangle \,.\label{exvalsto} \eea The matrix element   is evaluated in appendix (\ref{app:corres}), it is given by eqn. (\ref{adagera}). Using the symmetry property under exchange of the amplitudes, namely $\vk_1,\lambda_1 \Leftrightarrow \vk_2,\lambda_2$,    and upon relabelling the sum over the summed-over free polarization index in two of the terms, we find
 \be \langle \psi_{\gamma \gamma}(t)| a^\dagger_{{\vp}\,\alpha'}\,a_{\vp\, \alpha}   | \psi_{\gamma \gamma}(t)\rangle = \sum_{\lambda_2} C^*_{\gamma\gamma}(\vp,\alpha';\vp_2,\lambda_2;t) C_{\gamma\gamma}(\vp,\alpha; \vp_2,\lambda_2;t)~~;~~  {\vp}_2 = \vk-\vp\,. \label{finmtx}\ee By using the result (\ref{kinpol}) the matrix elements $\mathcal{J}_{\alpha'\alpha}(\vp;t)$ simplify to
 \be \mathcal{J}_{\alpha'\alpha}(\vp;t) = |\mathcal{K}(p,p_2;t)|^2\,\sum_{\lambda_2=1}^2 \mathcal{P}(\vp,\alpha',\vp_2;\lambda_2)\mathcal{P}(\vp,\alpha,\vp_2;\lambda_2)~~;~~\vp_2=\vk-\vp\,.\label{stokmel}\ee
 Using the general identities (\ref{polacorregen},\ref{unasuma}), we finally find ($\vp_2=\vk-\vp$)
 \be \mathcal{J}_{11}(\vp;t) = \mathcal{J}_{22}(\vp;t) =  |\mathcal{K}(p,p_2;t)|^2\,  \Big(1-\hat{\vp}\cdot\hat{\vp}_2 \Big)^2 ~~;~~  \mathcal{J}_{12}(\vp;t)= \mathcal{J}_{21}(\vp;t)=0\,,\label{Jsfin} \ee from which we infer that the degree of polarization (\ref{depol})
 \be \mathcal{D}(\vp;t) =0\,,\label{dipo}\ee concluding that in \emph{average} the two-photon state $|\psi_{\gamma \gamma}(t)\rangle$ is unpolarized. An important bonus of the result (\ref{S0}) is that it makes contact with the total photon number: it follows from eqn. (\ref{S0}) that the photon occupation number
 \be n_\gamma(\vp;t) = 2 \,\mathcal{S}_0(\vp;t) = 2\,\big( \mathcal{J}_{11}(\vp;t) + \mathcal{J}_{22}(\vp;t) \big) \,.\label{photnu}\ee Combining the above results with those of
 appendix (\ref{app:alltime}), we find
 \be N_\gamma(t) = \sum_{\vp} n_\gamma(\vp;t) = 2\,\Big(1-e^{-\Gamma t}\Big) \,,\label{totafo}\ee as expected from the decay of the axion quasiparticle into two photons.

 As stated above, at the quantum level the degree of polarization, involving solely the expectation values of the Stokes operators is not sufficient to characterize the polarization properties of the two photon state, because correlations of the Stokes operators provide more information\cite{korolkova,aga}. Rather than obtaining the momentum-polarization correlations of the Stokes operators, we obtain the normal ordered\footnote{Since we are considering correlations between operators with different momenta normal ordering does not require a subtraction.}\cite{aga}  correlation function of the building blocks of these operators, namely the operator products $a^\dagger_{s_a}(\vp) \,a_{s_b}(\vp)$, with $s_{a,b}=1,2$. Using the results of appendix (\ref{app:corres}), we find
 \bea & & \langle \psi_{\gamma\gamma}(t)|a^\dagger_{s_1}(\vp_1)\,a^\dagger_{s_2}(\vp_2)\,a_{s_3}(\vp_1)\,a_{s_4}(\vp_2)|\psi_{\gamma\gamma}(t)\rangle   =  C^*_{\gamma \gamma}(\vp_1,s_1,\vp_2,s_2;t) C_{\gamma \gamma}(\vp_1,s_3,\vp_2,s_4;t) \nonumber \\ &  = & |\mathcal{K}(\vp_1,\vp_2;t)|^2\, \mathcal{P}(\vp_1,s_1;\vp_2,s_2)\,\mathcal{P}(\vp_1,s_3;\vp_2,s_4)  \,,\label{stokecorre}\eea for $\vp_2 = \vk-\vp_1$.

  In particular,  for $s_3=s_1;s_4=s_2$, it follows that
 \be  \langle \psi_{\gamma\gamma}(t)|a^\dagger_{s_1}(\vp_1)\,a^\dagger_{s_2}(\vp_2)\,a_{s_1}(\vp_1)\,a_{s_2}(\vp_2)|\psi_{\gamma\gamma}(t)\rangle   =  \mathbb{P}_{\gamma \gamma}(\vp_1,s_1;\vp_2,s_2;t)~~;~~\vp_2=\vk-\vp_1\,,\label{corrprob}\ee where $\mathbb{P}_{\gamma \gamma}(\vp_1,s_1;\vp_2,s_2;t)$ is the probability of coincident measurement of two-photons of momenta $\vp_1,\vp_2$ and polarizations $s_1,s_2$ respectively,  given by equation (\ref{2fotprob1}).  This identification allows us to relate the momentum correlation functions of the  diagonal Stokes operators $S_0,S_1$ straightforwardly as linear combinations of the two-photon probabilities,

 \bea \langle \psi_{\gamma\gamma}(t)|\mathcal{S}_0(\vp_1)\mathcal{S}_0(\vp_2)|\psi_{\gamma\gamma}(t)\rangle  & =  & \mathbb{P}_{\gamma \gamma}(\vp_1,1;\vp_2,1;t)+\mathbb{P}_{\gamma \gamma}(\vp_1,2;\vp_2,2;t)+\mathbb{P}_{\gamma \gamma}(\vp_1,1;\vp_2,2;t)+\mathbb{P}_{\gamma \gamma}(\vp_1,2;\vp_2,1;t)\nonumber \\ &  =  & \frac{1}{2}\,|\mathcal{K}(p_1,p_2;t)|^2\,  (1-\vp_1\cdot \vp_2)^2 \,.\label{S2corr0}\eea and the momentum correlation of the \emph{polarization asymmetry} is given by

 \bea  \langle \psi_{\gamma\gamma}(t)|\mathcal{S}_1(\vp_1)\mathcal{S}_1(\vp_2)|\psi_{\gamma\gamma}(t)\rangle  &  = &  \mathbb{P}_{\gamma \gamma}(\vp_1,1;\vp_2,1;t)+\mathbb{P}_{\gamma \gamma}(\vp_1,2;\vp_2,2;t)-\mathbb{P}_{\gamma \gamma}(\vp_1,1;\vp_2,2;t)-\mathbb{P}_{\gamma \gamma}(\vp_1,2;\vp_2,1;t)\nonumber \\ &  =  & - \frac{1}{2}\,\Big[\big(\hat{\vec{\epsilon}}_2(\vp_1)\cdot \hat{\vec{\epsilon}}_2(\vp_2) \big)^2-\big(\hat{\vec{\epsilon}}_1(\vp_1)\cdot \hat{\vec{\epsilon}}_2(\vp_2) \big)^2  \Big]\,|\mathcal{K}(p_1,p_2;t)|^2\,  (1-\vp_1\cdot \vp_2)^2~ \,.\label{S2corr1}\eea for $\vp_2=\vk-\vp_1$. Therefore, it is clear that even when the average of the polarization asymmetry operator vanishes, its fluctuations or momentum correlations do not, this is a manifestation of the momentum-polarization hyperentanglement. The result (\ref{S2corr1}) is of experimental relevance: by coincident measurements of the polarization of the two-photons yielding the probabilities $\mathbb{P}_{\gamma\gamma}$ for the various polarization configurations the momentum correlation of the polarization asymmetry may be extracted. Such measurement probes directly the distinct polarization pattern arising from parity and time reversal breaking from the Chern-Simons interaction (axion electrodynamics).

 For the off-diagonal Stokes operators, we find, for example

 \be \langle \psi_{\gamma\gamma}(t)|\mathcal{S}_2(\vp_1)\mathcal{S}_2(\vp_2)|\psi_{\gamma\gamma}(t)\rangle = \frac{1}{2}\Big[\big(\hat{\vec{\epsilon}}_2(\vp_1)\cdot \hat{\vec{\epsilon}}_2(\vp_2) \big)^2-\big(\hat{\vec{\epsilon}}_1(\vp_1)\cdot \hat{\vec{\epsilon}}_2(\vp_2) \big)^2  \Big]\, |\mathcal{K}(p_1,p_2;t)|^2\,(1-\vp_1\cdot \vp_2)^2\,,\label{S2corre}\ee

\be \langle \psi_{\gamma\gamma}(t)|\mathcal{S}_3(\vp_1)\mathcal{S}_3(\vp_2)|\psi_{\gamma\gamma}(t)\rangle = \frac{1}{2}\, |\mathcal{K}(p_1,p_2;t)|^2\,(1-\vp_1\cdot \vp_2)^2 \,,\label{S3corre}\ee for $\vp_2=\vk-\vp_1$, which clearly display correlation information even when
\be \langle \psi_{\gamma\gamma}(t)|\mathcal{S}_{i}(\vp) |\psi_{\gamma\gamma}(t)\rangle = 0 ~~;~~ i=1,2,3\,.\label{zeroS12}\ee The general result (\ref{stokecorre}) implies that the \emph{Stokes correlation matrix}
\be \langle \psi_{\gamma\gamma}(t)|\mathcal{S}_i(\vp_1)\mathcal{S}_j(\vp_2)|\psi_{\gamma\gamma}(t)\rangle = \frac{1}{2}\, |\mathcal{K}(p_1,p_2;t)|^2\,(1-\vp_1\cdot \vp_2)^2\,\mathbb{V}_{ij}(\varphi)\,,\label{stokmtx}\ee where the $3\times 3$ correlation matrix $\mathbb{V}_{ij}(\varphi)$ is solely a function of the angle $\varphi$.

\subsection{Hanbury-Brown Twiss (HBT) second order coherence:}\label{subsec:hbt}
The characteristics of the two photon state may be probed with intensity interferometry as per Glauber's theory of photodetection\cite{glauber} with the
Hanbury-Brown Twiss\cite{hbt} (HBT) second order coherence for coincident photo detection
\be G^{(2)}(\vec{x}_1,t;\vec{x}_2,t) = \bra{\psi(t)}E^{-}(\vec{x}_2,t)E^{-}(\vec{x}_1,t)E^{+}(\vec{x}_1,t)E^{+}(\vec{x}_2,t)\ket{\psi(t)} \,,\label{G2}\ee (we suppressed the vector labels) where
\bea \vec{E}^{-}(\vec{x},t)  & = &  {-i} \sum_{\vk,\lambda=1,2} \sqrt{\frac{k}{2V}}\, \vec{\epsilon}_{\lambda}(\vk)\,a^\dagger_\lambda(\vk)\,e^{i(\vk\cdot \vx-kt)}\,,\label{Emin}\\
\vec{E}^{+}(\vec{x},t)  & = &  {i} \sum_{\vk,\lambda=1,2} \sqrt{\frac{k}{2V}}\, \vec{\epsilon}_{\lambda}(\vk)\,a_\lambda(\vk)\,e^{-i(\vk\cdot \vx-kt)}\,. \label{Eplus}\eea
As discussed above, the two photon state features correlations both in momentum and polarization which are a distinct telltale of the parity and time reversal breaking interaction of axion electrodynamics. However,  the operators (\ref{Emin},\ref{Eplus}) are   sums of both observables, hence, the spatio-temporal second order coherence (\ref{G2}) probes momentum and polarization entanglement only indirectly. See further discussion in section (\ref{sec:discussion}).
To access momentum and polarization correlations with second order (HBT) coherences, photodetection must be performed with narrowband frequency (momentum) filters
set up in conjuction with polarizers.

Therefore, following ref.\cite{mandl} we introduce instead the momentum and polarization resolved operators
\bea \mathcal{E}^{-}_s(\vec{p};t) & = &  {\sqrt{\frac{1}{V}}}\int   \vec{\epsilon}_s(\vec{p})\cdot \vec{E}^{-}(\vx,t) \,e^{-i \vec{p}\cdot \vx}  \, d^3 x \,,\label{Omin}\\\mathcal{E}^{+}_s(\vec{p};t) & = &  {\sqrt{\frac{1}{V}}}\int  \vec{\epsilon}_s(\vec{p})\cdot \vec{E}^{+}(\vx,t) \,e^{i \vec{p}\cdot \vx} \, d^3 x\,,\label{Oplus} \eea which are directly related to the photon annihilation and creation operators as
\be \mathcal{E}^{-}_s(\vec{p};t) = \sqrt{\frac{p}{2}}\,a_s(\vec{p})\,e^{-ipt}~~;~~ \mathcal{E}^{+}_s(\vec{p};t) = \sqrt{\frac{p}{2}}\,a^\dagger_s(\vec{p})\,e^{ipt} ~~;~~ s=1,2\,. \label{Oarel}\ee

Hanbury-Brown Twiss interferometry with frequency resolved (``colored'') photon correlations has been experimentally implemented  in ref.\cite{hbtcolor}, providing direct confirmation    of the feasibility of studying frequency (momentum) filtered second order coherence. Motivated by this implementation,   instead of the spatio-temporal (HBT) correlation (\ref{G2}), and following ref.\cite{mandl} we introduce the \emph{momentum and polarization} resolved cross-correlation (second order cross-spectral density)\footnote{See chapter 12.5.2 in ref.\cite{mandl}.}

\be \widetilde{G}^{(2)}(\vec{p}_1,s_1;\vec{p}_2,s_2;\vec{p}_3,s_3;\vec{p}_4,s_4;t) =  \bra{\psi(t)}a^{\dagger}_{s_1}(\vec{p}_1) a^{\dagger}_{s_2}(\vec{p}_2) a_{s_3}(\vec{p}_3)  a_{s_4}(\vec{p}_4) \ket{\psi(t)} \,,\label{tilG2}\ee  from which the full spatio-temporal second order coherence (\ref{G2})  may be obtained.
Using the symmetry $C_{\gamma \gamma}(\vec{p}_1,s_1;\vec{p}_2,s_2;t)= C_{\gamma \gamma}(\vec{p}_2,s_2;\vec{p}_1,s_1;t)$ we find
\be \widetilde{G}^{(2)}(\vec{p}_1,s_1;\vec{p}_2,s_2;\vec{p}_3,s_3;\vec{p}_4,s_4;t) = C^*_{\gamma \gamma}(\vec{p}_1,s_1;\vec{p}_2,s_2;t)\,C_{\gamma \gamma}(\vec{p}_3,s_3;\vec{p}_4,s_4;t)~~;~~\vp_2=\vk-\vp_1;\vp_4=\vk-\vp_3 \,.\label{tilGfin} \ee

 Although ref.\cite{hbtcolor} did not consider the spatial dependence of the electric fields, the momentum correlations  (\ref{tilG2}) should be interpreted as a narrow-band limit of the frequency resolved correlations introduced in  ref.\cite{hbtcolor},  suggesting that the may also be studied with a similar framework.

Furthermore, there is a close relationship between the momentum and polarization resolved second order correlation  (\ref{tilG2}) and the momentum correlation of the Stokes operators. Indeed, the particular case $\vp_3 = \vp_1;\vp_4=\vp_2$  of the second order correlation (\ref{tilG2}) can be immediately identified with the correlations (\ref{stokecorre}) which are the building blocks of the momentum correlations of the Stokes operators as implemented above.

The main conclusions of this analysis are that whereas the average degree of polarization and polarization asymmetry in the two-photon quantum state arising from the
time evolution of dynamical axion decay vanish, momentum-polarization hyperentanglement is imprinted in momentum correlations of the quantum Stokes operators. The  distinct polarization pattern is a consequence of the parity and time reversal breaking
interaction of dynamical axions and photons. In particular the momentum correlation of the polarization asymmetry, given by equation (\ref{S2corr1}),  can be directly assessed from the probability of coincident two-photon detection for various configurations of the polarizer  .

\subsection{Similarities and differences with parametric down conversion:}\label{subsec:spdc}
The results in the previous sections are strikingly similar to the phenomenon of parametric down conversion\cite{pdc1,pdc2,pdc3}. In this process   an incoming
beam of photons from a laser (pump) propagate in a non-linear medium and split into two lower frequency signal and idler photons through a non-linear second order susceptibility tensor $\chi_{ijk}$. The interaction Hamiltonian is modeled with a cubic type-electromagnetic vertex  $\propto \chi_{ijk}E^P_{i}E^S_jE^I_k$, with $P,S,I$ for pump, signal and idler fields\cite{pdc1,pdc2,pdc3}. For an intense monochromatic laser beam, the pump field is usually taken to be classical whereas the signal and idler fields are quantized. The two photons emerging from the non-linear crystal (signal and idler) feature momentum and polarization entanglement as a consequence of the non-linearity similarly to the two photons emerging from axion decay. The down conversion is of type I or type II respectively if the photon pair features parallel or perpendicular polarizations.

In a finite sized crystal and during a finite time interval,  the momenta and frequencies of the signal and idler photons are determined by the \emph{sinc}-type functions\cite{pdc1}
\be \Pi^3_{m=1}\,\Bigg[ \frac{\sin\big[\frac{1}{2}\big(\vec{k}_P-\vec{k}_S-\vec{k}_I \big)_m L_m \big]}{\frac{1}{2}\big(\vec{k}_P-\vec{k}_S-\vec{k}_I \big)_m}\Bigg] \times \frac{\sin\big[\frac{1}{2}\big(E_P-k_S-k_I \big) t \big]}{\frac{1}{2}\big(E_P-k_S-k_I \big) } \,, \label{sincs} \ee which for large size in each direction $L_m$ and for large time $t$ yield the (approximate) phase and frequency matching conditions
\be \vec{k}_P \simeq  \vec{k}_S+\vec{k}_I ~~;~~ E_P\simeq k_S+k_I \,,\label{match}\ee with $\vk_P;E_P$ being the momentum and energy of the pump field. These conditions must be compared to the momentum conservation condition in axion decay $\vec{k}_1+\vec{k}_2 = \vec{k}$ in the results of the previous section and the resonant denominator in the amplitude (\ref{kinem})   yielding the largest amplitude for the energy conserving conditions $E_{2\gamma} = k_1+k_2 \simeq E_{\phi}$. In the limit of vanishing width the resonant denominator in the two photon probability (\ref{2fotprob2})  can be replaced by approximate energy conserving delta functions, the ``blurring'' of the energy conserving condition is a consequence of the uncertainty associated with the lifetime of the decaying state. The time dependent \emph{sinc} function in (\ref{sincs}) is the usual function arising in time dependent perturbation theory that yields energy conservation in the long time limit a la Fermi's golden rule. Therefore the kinematic constraints from parametric down conversion (phase and frequency matching) are equivalent to those in the two-photon final state
from axion decay, this is of course expected from the kinematics of energy-momentum conservation. The main difference is in the polarization entanglement. Whereas for axion decay, the polarization function  is given by eqn. (\ref{polac}) and is direct consequence of the parity and time reversal breaking in axion electrodynamics, in parametric down conversion the dependence in the polarization is given by a function of the form\cite{pdc1}
\be \mathcal{C}(\vec{k}_S,\lambda_S;\vec{k}_I,\lambda_I) \propto \widetilde{\chi}_{ijl}\,\vec{\epsilon}_j(\vec{k}_S) \,\vec{\epsilon}_l(\vec{k}_I) \,,\label{cpolapdc}\ee where $\widetilde{\chi}$ is a frequency Fourier transform of the non-linear susceptibility.

These differences notwithstanding, the main remarkable similarity between axion decay and parametric down conversion is that of momentum and polarization hyperentanglement in the final two photon state. In references\cite{pdc1,pdc2,pdc3} second order (HBT) (intensity) interferometry,   has been proposed to study the correlations between signal and idler photons, in either type of parametric down conversion. And in ref.\cite{bi,kiess,shih} polarization resolved (HBT) second order coherence was used to test violations of Bell inequalities in a two-photon correlation experiment from the interference of signal and idler photons from parametric down conversion.

More recently alternative sources of entangled photons have been studied such as  excitons and bi-excitons\cite{exci,tur}, semiconductor quantum dots\cite{dousse,akopian,kwiat1,kwiat2} and interacting quantum emitters\cite{delgado}. The similarities between the two-photon quantum state in these systems and that of dynamical axion decay  suggest that the experimental  techniques   implemented in these references may be harnessed to probe dynamical axion quasiparticles with   two-photon correlations, complementing the proposal in ref.\cite{olivia} for multiphoton detection of axionic quasiparticles.

\section{Discussion}\label{sec:discussion}

\begin{figure}[ht]
\includegraphics[height=2.5in,width=2.5in,keepaspectratio=true]{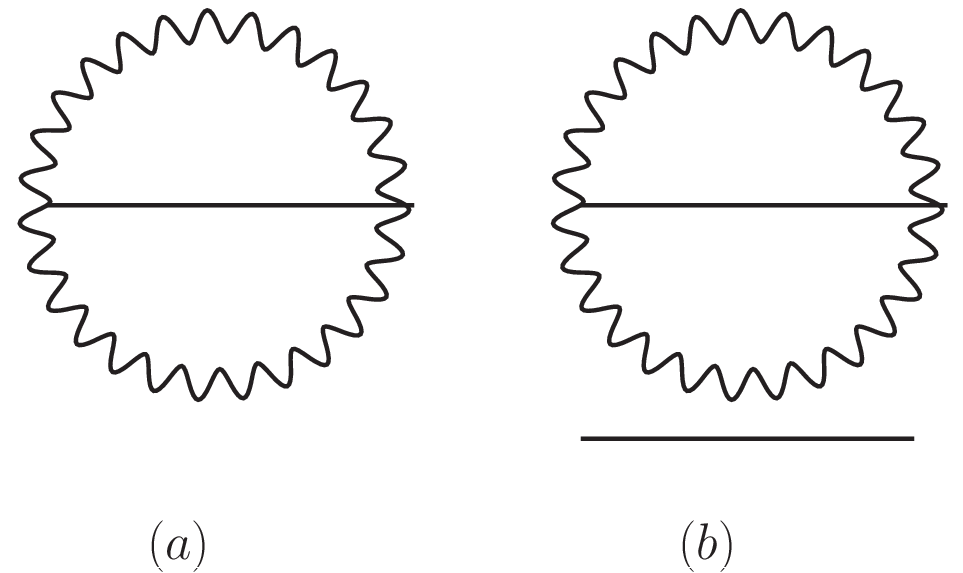}
\caption{Non (RWA) contributions: fig. (a) is a vacuum diagram, fig. (b) is a disconnected diagram with the single axion quasiparticle and vacuum correction. Wiggly lines are photon lines, solid lines correspond to a single axion particle. These diagrams yield a   phase $e^{i\Delta E_0 t}$, with $\Delta E_0$ the renormalization of the vacuum energy.   }
\label{fig:vacuum}
\end{figure}

\textbf{Non (RWA) contributions:} In the interaction Hamiltonian (\ref{HIcs}) we have kept only the terms that yield resonant (energy conserving) contributions, namely the (RWA). The remaining contributions, denoted by $(\cdots)$ within the bracket in eqn. (\ref{HIcs}) are
\be  \Bigg[ b^\dagger(\vp) \, a^\dagger_{\lambda_1}(\vp_1) a^\dagger_{\lambda_2}(\vp_2) \,e^{i(E_\phi(p)+p_1+p_2)t} +  b(\vp) \, a_{\lambda_1}(\vp_1) a_{\lambda_2}(\vp_2) \,e^{-i(E_\phi(p)+p_1+p_2)t}\Bigg] \,,\label{nonrwa} \ee the first term creates and the second annihilates an axion and two photons. Up to second order, considered in this study, these contributions yield disconnected vacuum diagrams: consider that the initial state is simply the vacuum for all fields, the final state (in second order) is again the vacuum with a three particle intermediate state, a vacuum disconnected diagram displayed in fig. (\ref{fig:vacuum}) (a). For an initial single axion particle state, the contributions from (\ref{nonrwa}) yield the disconnected diagram (\ref{fig:vacuum}) (b). Following the steps in section (\ref{sec:ww}), it is straightforward to show that the disconnected diagrams yield a  phase $e^{i\Delta E_0 t}$ with $\Delta E_0$ the vacuum energy shift (renormalization). Just as in S-matrix theory, the disconnected diagrams represent a renormalization of the vacuum energy and is common to all diagrams with arbitrary number of particles in the initial state.

Since we focus on the final two-photon state arising from the decay of the axion quasiparticle, we neglect these disconnected vacuum contributions which are absorbed into a vacuum redefinition.

\textbf{Spatio-temporal correlations:} As shown in the previous sections, the two-photon quantum state from axion decay is entangled both in momentum and polarization. Therefore, we focused on studying correlations of appropriate operators in these physical degrees of freedom, identifying the quantum Stokes operators and relating them to momentum and polarization resolved (HBT) second order coherences. However, it is  appropriate to ask whether the usual spatio-temporal (HBT) correlations (\ref{G2}) yield complementary information on the two-photon state such as, for example, if there is any  imprint of entanglement and the distinct pattern of polarization from the parity and time reversal breaking interaction in the phenomenon of photon bunching.  There are several aspects that must be clearly understood to answer these questions: \textbf{i:)} we have purposely neglected the vector labels of the electric field operators in (\ref{G2}), each one relating to a three dimensional vector. In Glauber's\cite{glauber,zubairy,meystre} seminal theory of photodetection these indices are contracted with those of the atomic dipole moments, therefore,  they are ``free indices'' in (\ref{G2}) which in the field expansion are carried by the polarization vectors. Hence, a first question to address is how to treat these vector indices. \textbf{ii:)} in the frequency filtered ``colored'' (HBT) in ref.\cite{hbtcolor} neither the spatial dependence nor the vector labels of the electric field   were included in the analysis, which only studied the temporal correlations, therefore the polarization degree of freedom was not explicitly included. \textbf{iii:)} The spatio-temporal correlations imply an integral over the wave vectors and sum over polarizations for each one of the electric field components thereby ``blurring'' the distinct pattern of entanglement in the two-photon quantum state. Photon bunching in parametric down conversion was studied in ref.\cite{ouwang}, however, neither the spatial dependence (momentum) nor polarization vectors are explicitly introduced in the analysis.   \textbf{iv:)} (HBT) correlations of Stokes parameters were studied in refs.\cite{stokes2,stokes3}, but the treatment only considered beams (typically Gaussian) propagating along one axis (unidirectional) with a well defined polarization plane perpendicular to the propagation axis. This is far from the case of the quantum two photon state from axion decay where there is no preferred directionality of emission and polarization plane. Therefore, the question of how to tease out momentum-polarization entanglement with the distinct pattern from Chern simons coupling from spatio-temporal HBT correlations remains to be understood and will be the subject of further study elsewhere.

\textbf{Other decay channels.} Since we are focused on probing dynamical axion quasiparticles with photon correlations, only the ``visible'' decay channel into photons is relevant for this study. The axion-photon interaction (\ref{synacfin}) determines that axions only decay into an even number of photons, the ``tree level'' contribution to the decay rate is given by eqn. (\ref{Gamafi}) and from the optical theorem (unitarity) it corresponds to the imaginary part of the self energy on the axion mass shell. The next available decay channel is a 4-photon final state, this contribution features both an axion and a photon internal propagators, its amplitude is $\propto g^3$ yielding a decay rate into 4 photons $\propto g^6$, being the imaginary part of a \emph{three loop} self-energy. This channel is suppressed with respect to the leading order (2-photons) by a large power of the (small) coupling $g^4$.

An important question is on the coupling of the dynamical axion field to \emph{electron-hole pairs}: the Chern-Simons coupling of the dynamical axion to electromagnetic fields is
a consequence of the triangle (chiral) anomaly,  which emerges from a direct coupling of the dynamical axion field to electronic degrees of freedom, which in turn couple to electromagnetic fields. The triangle diagram is a one loop contribution that features an electron-hole pair in the intermediate state, therefore it is an important question of
whether the direct coupling of the dynamical axion  to electron-hole pairs can provide an alternative probe of the axion field. Although such processes may be suppressed by gap effects in the bulk or surface effects, we cannot assess this important question within the framework of the effective quantum field theory of the emergent dynamical axion field defined by the Lagrangian (\ref{synac}). This is because to obtain this effective action the electronic degrees of freedom (responsible for the triangle diagram) have been integrated out\cite{xiao,rundong,jing,nomura}. The effective action (\ref{synac}) has no information on these electronic degrees of freedom, preventing us from even inquiring about how the direct coupling of the dynamical axion to electron-hole pairs may yield alternative probes. Furthermore, as emphasized in the introduction, we are focused on understanding the ``visible'' (by this we mean photons) decay channels to study the distinct pattern of photon correlations (momentum and polarization) as observables. However, further studies including the underlying electronic degrees of freedom merit detailed consideration, which is beyond the scope and framework of this article.

Although a hallmark of dynamical axion quasiparticles is their interaction with photons via the parity and time reversal breaking
Chern-Simons term $\propto \vec{E}\cdot \vec{B}$, collective excitations in topological insulators and in Weyl semimetals may also couple to other non-gauge degrees of freedom resulting in other non-electromagnetic decay channels. Unlike the ``universal'' form of the coupling of dynamical axions to electromagnetic fields, interaction with other degrees of freedom may depend on the details of the material. These interactions may affect the consequences of radiative decay into photons via renormalization of the axion-photon vertex via intermediate states in various ways which must be studied in detail in specific cases but that will be suppressed with respect to the main decay channel by powers of the axion couplings to other degrees of freedom. These may affect the effective vertex but  will \emph{not contribute} to the momentum-polarization photon correlations, studied here, since these depend solely on the pseudoscalar coupling to $\vec{E}\cdot \vec{B}$.   Undoubtedly such possibilities merits further and deeper study, which, however, is well beyond the scope of this article.

\textbf{Resolution aspects:} There are (at least) two sources of resolution uncertainties that could affect the observability of two-photon correlations: \textbf{i:)} lifetime broadening of the spectral distribution, \textbf{ii:)} experimental resolution. The lifetime broadening is completely determined by the factor $\mathcal{K}(k_2,k_2;t)$, given by equation (\ref{kinem}) and determines the Lorentzian distribution in the two-photon probability (\ref{2fotprob2}) which is very narrow for small axion-photon coupling. However, we highlight that the telltale polarization dependence is completely determined by the function $\mathcal{P}(\vk_1,\lambda_1;\vk_2,\lambda_2)$ given by equation (\ref{polac}) which multiplies the Lorentzian spectral density, therefore although the momentum resolution is broadened by the lifetime, the \emph{polarization correlation} along with momentum-polarization hyperentanglement are not affected by this time-energy uncertainty.

The experimental limitations on resolution are, obviously, aspects that must be addressed within a particular experimental setting, however, the experimental setups and results exploiting momentum filtering\cite{hbtcolor} and momentum and polarization hyperentanglement in
type II parametric down conversion reported in references[45-56] with similar time-energy uncertainty broadening given by equation (\ref{sincs}), suggest that the experimental resolution aspects can be systematically controlled without modifying the two-photon correlations and the observational consequences of momentum-polarization hyperentanglement.

\section{Conclusions and further questions}\label{sec:conclusions}
Dynamical axion quasiparticles, collective excitations in topological insulators with broken parity and time reversal invariance, or charge density waves in Weyl semimetals are currently the focus of intense theoretical and experimental studies. They bear many characteristics of  the axion field in particle physics, a hypothetical particle beyond the Standard Model and compelling dark matter candidate, and could also provide novel platforms for quantum information and quantum computing. Recently, observation of coherent oscillations induced by antiferromagnetic magnons in two dimensional ($MnBi_2Te_4$) was interpreted as evidence for  dynamical axion quasiparticles, bolstering the case for their deeper study. A hallmark of axion degrees of freedom is their coupling to photons via a topological parity and time reversal breaking Chern-Simons term $\propto \vec{E}\cdot \vec{B}$, which implies a decay channel into two photons.

Our objective in this study is to focus on this process, to obtain the two-photon state from axion decay and to study its properties, in particular polarization aspects, to identify possible observational signatures that could be harnessed to probe dynamical axions with two-photon correlations. To achieve this aim we extended the Weisskopf-Wigner theory of atomic spontaneous decay to the effective field theory of dynamical axions, allowing us to obtain the quantum two-photon state from its decay. The particular form of axion-photon interaction yields a final two-photon state with a distinct pattern of of momentum and polarization hyperentanglement. We introduce quantum Stokes operators to quantify the degree of polarization, polarization asymmetry and correlations  in the two-photon state. We find that whereas the  \emph{average} of the degree of polarization and polarization asymmetry vanish in the two-photon state, there are non-trivial momentum correlations of the Stokes operators as a consequence of momentum-polarization entanglement. These correlations are a manifestation of the polarization pattern resulting from the particular interaction Hamiltonian and  can be  revealed with coincident two-photon detection.

We obtain a direct relationship between the momentum correlation of Stokes operators and momentum and polarization resolved Hanbury-Brown Twiss second order coherence. There is a striking similarity between dynamical axion decay into two correlated photons and spontaneous parametric down conversion in non-linear (Kerr) optical media as well as other  condensed matter matter systems featuring spontaneous decay into entangled photon pairs. This similarity and the relationship between the momentum correlations of Stokes operators and Hanbury-Brown Twiss second order coherence  suggests that experimental setups designed to test Bell inequalities with two-photon correlations from parametric down conversion may be usefully implemented to probe the emergence of dynamical axion quasiparticles with two-photon correlations.

\textbf{Further questions:} In this study we focused on obtaining correlations of Stokes operators in momentum space since these directly probe the momentum and polarization entanglement of the two photon state from dynamical axion decay, relating them to particular configurations of momentum and polarization resolved second order coherence.  A question that remains to be addressed is if and how momentum-polarization entanglement is imprinted in \emph{spatio-temporal} Hanbury-Brown Twiss intensity correlations, for example whether there is any manifestation of these correlations in the phenomenon of photon bunching in space and or time. As these spatio-temporal correlations entail integration over momenta, the polarization components that depend on the momenta are also integrated. We recognize that in order to understand these issues, it remains to clarify how the vector nature of the (normal ordered) field operators in the second order coherence is treated in these correlations, an aspect   that while clearly related to the issue of photon correlations is relegated to further study.

\acknowledgements
 The author    gratefully acknowledges  support from the U.S. National Science Foundation through grants   NSF 2111743 and NSF 2412374.

\appendix

\section{Evaluation of $\rho(k_0)$:}\label{app:spec}
The evaluation of the spectral density requires $\sum_{\lambda_1,\lambda_2}\Big[ \mathcal{P}(\vk_1,\lambda_1;\vk_2,\lambda_2)  \Big]^2$ with $\mathcal{P}(\vk_1,\lambda_1;\vk_2,\lambda_2) $ the polarization function  (\ref{polac}). There are altogether four terms, which are of two types

\textbf{Type I:}

\bea &  &  \sum_{\lambda_1} {\epsilon}^{i}_{\lambda_1}(\vk_1) {\epsilon}^{j}_{\lambda_1}(\vk_1) \times \sum_{\lambda_2}
\big(\hat{\vk}_2\times \hat{\vec{\epsilon}}_{\lambda_2}(\vk_2) \big)^{i}\big(\hat{\vk}_2\times \hat{\vec{\epsilon}}_{\lambda_2}(\vk_2) \big)^{j} + (\vk_1 \Leftrightarrow \vk_2)\\ & = & 2\,\big[\delta^{ij}- \hat{\vk}^i_1\,\hat{\vk}^j_1 \big]\,\big[\delta^{ij}- \hat{\vk}^i_2\,\hat{\vk}^j_2 \big] = 2\big[1+(\hat{\vk}_1\cdot\hat{\vk}_2)^2\big]\,\label{type1} \eea

\textbf{Type II:}
\be 2\times \sum_{\lambda_1} {\epsilon}^{i}_{\lambda_1}\big(\hat{\vk}_1\times \hat{\vec{\epsilon}}_{\lambda_1}(\vk_1) \big)^{j}\,\sum_{\lambda_2} \big(\hat{\vk}_2\times \hat{\vec{\epsilon}}_{\lambda_2}(\vk_2) \big)^{i}\,{\epsilon}^{j}_{\lambda_2}\big(\hat{\vk}_2\big) = -4\,\hat{\vk}_1\cdot \hat{\vk}_2 \,.\label{type2} \ee This result is proven by using the relations (\ref{polas}) from which we find
\bea &&  \sum^{2}_{\lambda_1=1} {\epsilon}^{i}_{\lambda_1}\big(\hat{\vk}_1\big) \big(\hat{\vk}_1\times \hat{\vec{\epsilon}}_{\lambda_1}(\vk_1) \big)^{j} = {\epsilon}^{i}_{1}\big(\hat{\vk}_1\big) \,{\epsilon}^{j}_{2}\big(\hat{\vk}_1\big) -{\epsilon}^{j}_{1}\big(\hat{\vk}_1\big) \,{\epsilon}^{i}_{2}\big(\hat{\vk}_1\big) \nonumber \\ &&
\sum^{2}_{\lambda_2=1} \big(\hat{\vk}_2\times \hat{\vec{\epsilon}}_{\lambda_2}(\vk_2) \big)^{i}\,{\epsilon}^{j}_{\lambda_2}\big(\hat{\vk}_2\big) = -\Big( {\epsilon}^{i}_{1}\big(\hat{\vk}_2\big) \,{\epsilon}^{j}_{2}\big(\hat{\vk}_2\big) -{\epsilon}^{j}_{1}\big(\hat{\vk}_2\big) \,{\epsilon}^{i}_{2}\big(\hat{\vk}_2\big) \Big)\,,\label{type2a} \eea the remaining scalar products are obtained using the relations (\ref{k2vecs}) yielding the result (\ref{type2}). Gathering (\ref{type1}), and (\ref{type2}) we finally obtain
\be \sum_{\lambda_1,\lambda_2}\Big[ \mathcal{P}(\vk_1,\lambda_1;\vk_2,\lambda_2)  \Big]^2 = 2\,\big[1- \hat{\vk}_1\cdot \hat{\vk}_2\big]^2\,.\label{Psquared}  \ee

With $\vk_2=\vk-\vk_1$ and
\be \vk\cdot \vec{k}_1 = k k_1\cos(\alpha)~~;~~ d^3 k_1 =  2\pi\, k^2_1 dk_1 d(\cos(\alpha))\,\label{dotprod}\ee and defining
\be q\equiv |\vec{k}-\vec{k}_1| = \sqrt{k^2+k^2_1 - 2kk_1\cos(\alpha)}  \Rightarrow \frac{d(\cos(\alpha))}{|\vk-\vk_1|} =- \frac{dq}{k k_1} \,\label{qdef}\ee the spectral density (\ref{rhoaxfin}) becomes
\be \rho (k_0) = \frac{g^2}{64\pi^2 k \, E_{\phi}(k)}\,\int^\infty_0 dk_1 \int^{q^+}_{q^-} \Big[(q+k_1)^2-k^2 \Big]^2\,\delta(k_0-k_1-q) \,dq\,,\label{rhosi}\ee
where the upper and lower limits are given by
\be q^+ = (k_1+k) ~~;~~ q^- = |k-k_1| \,.\label{qpm}\ee Using the delta function   constraint, yielding  $q+k_1 = k_0$ leads to
\be \rho(k_0) = \frac{g^2}{64\pi^2 k \, E_{\phi}(k)}\,\Big(k^2_0-k^2\Big)^2\,\int^\infty_0 dk_1 \int^{q^+}_{q^-}  \delta(k_0-k_1-q) \,dq\,.\label{rhosita}\ee The q-integral is non-vanishing and  equal to one  if the conditions
\be  k_0 >0 ~~;~~ |k_1-k| \leq k_0-k_1 \leq k+k_1\,,\label{condis}\ee are fulfilled. A straightforward analysis shows that these   are fulfilled for
\be k_0 > k ~~;~~ k_- \leq k_1 \leq k_+  \ee  where
\be k_{\pm} = \frac{1}{2} (k_0 \pm k)\,, \label{kpm}\ee finally yielding the result (\ref{rhoabfinal}).

\section{Solution of eqn. (\ref{dotC1}) by Laplace transform:}\label{app:lapla}
With the spectral density (\ref{rhosd}) the amplitude equation (\ref{dotC1}) can be written as
\be \dot{C}_{\phi}(t)   =  -  \int^t_0  \int_{-\infty}^{\infty}\rho(k_0)\, e^{i(E_\phi-k_0)(t-t')} dk_0\,   {C}_{\phi}(t')   \, dt'\,, \label{dotCnew} \ee
Introducing the Laplace variable $s$ and the Laplace transform of $C_\phi(t)$ as $\widetilde{C}_\phi(s)$, with the initial condition $C_\phi(t=0)=1$, we find \be \widetilde{C}_\phi(s)= \Bigg[s+\int_{-\infty}^\infty dk_0 ~ \frac{\rho(k_0)}{s+i(k_0-E_\phi)}\Bigg]^{-1} \label{Lapla} \ee with solution \be C_\phi(t) = \int^{i\infty+\epsilon}_{-i\infty +\epsilon} \frac{ds}{2\pi i} ~\widetilde{C}_\phi(s)\,e^{st} \label{invlapla}\ee where the $\epsilon \rightarrow 0^+$ determines the Bromwich contour in the complex $s$-plane parallel to the imaginary axis to the right of all the singularities of $\widetilde{C}_\phi(s)$. Changing variables to  $s=i(\omega-i\epsilon)$ we find \be C_A(t) = \int_{-\infty}^{\infty} \frac{d\omega}{2\pi\,i}~ \frac{e^{i\omega t}}{\Bigg[\omega-i\epsilon - \int_{-\infty}^{\infty} dk_0~\frac{\rho(k_0)}{\omega+k_0-E_\phi-i\epsilon} \Bigg]  }\label{CAfin}\ee

In the free case where $\rho =0$, the pole is located at $\omega =i\epsilon$ with $\epsilon \rightarrow 0^+$, leading to a constant $C_\phi=1$. In perturbation theory there is a complex pole very near $\omega =0$, at $\omega \propto \mathcal{O}(H^2_I)$ which can be obtained directly by setting $\omega=0$ in the integral in the denominator.
  We find in the limit $\epsilon \rightarrow 0^+$ \be  \int_{-\infty}^{\infty} dk_0~\frac{\rho(k_0)}{ k_0-E_\phi-i\epsilon} = -\Delta  + i \,\frac{\Gamma}{2} \label{aproxi}\ee where
  \be  \Delta      =   \mathcal{P} \int_{-\infty}^{\infty} dk_0 \, \frac{\rho(k_0)}{(E_\phi-k_0)} ~~;~~\Gamma  =  2\pi\,\rho(E_\phi)     \ee and $\mathcal{P}$ stands for the principal part. The energy shift (Lamb shift) $\Delta  $   and decay rate $\Gamma $ are precisely those obtained in the Markov approximation  (\ref{RIsig}). The {\em long time} limit of $C_\phi(t)$ is determined by this complex pole near the origin in the complex $s$-plane, leading to the asymptotic behavior \be C_\phi(t)= \, e^{-i\Delta  t}\,e^{-\frac{\Gamma}{2}\,t} \label{tasi}\ee
  which agrees with the result (\ref{cfisol}) obtained in the Markov approximation.

  \subsection{Beyond the Markov approximation:}\label{subsec:nonmarkov}
  Equation (\ref{dotc1fin}) may also be solved exactly without invoking the long time limit (\ref{pp}), namely the Markov approximation. From the definitions
  (\ref{iden}) (including the indistinguishability factor for the case of the two photon final state), and (\ref{rhosd}) for the spectral density, it follows that
  \be W[t,t] = \int^t_0 \int^{\infty}_{-\infty}  dk_0 \, \rho(k_0) \, e^{i(E_{\phi}-k_0)(t-t')}\,dt' = -i\,\int^{\infty}_{-\infty} \rho(k_0) \Bigg[\frac{1-e^{-i(k_0-E_\phi)t}}{(k_0-E_\phi)}   \Bigg]\,dk_0 \,.\label{dobleW}\ee yielding the solution of (\ref{dotc1fin}) as
  \be C_\phi(t) = e^{-i\mathcal{E}(t)}\,,\label{curE}\ee with
  \be  \mathcal{E}(t)   =   t \, \int^{\infty}_{-\infty} \frac{\rho(k_0)}{(k_0-E_\phi)} \Bigg[1-\frac{\sin\big((k_0-E_\phi)\,t\big)}{(k_0-E_\phi)\,t}   \Bigg]\,dk_0
    -i    \int^{\infty}_{-\infty}  \frac{{\rho(k_0)} }{(k_0-E_\phi)^2} \Big[1- {\cos\big((k_0-E_\phi)\,t\big)}    \Big]\,dk_0 \,. \ee
    Asymptotically as $t \rightarrow \infty$ these integrals yield
    \bea \int^{\infty}_{-\infty} \frac{\rho(k_0)}{(k_0-E_\phi)} \Bigg[1-\frac{\sin\big((k_0-E_\phi)\,t\big)}{(k_0-E_\phi)\,t}   \Bigg]\,dk_0    & ~{}_{\overrightarrow{t\rightarrow \infty}} ~ &  \mathcal{P}  \int^{\infty}_{-\infty} \frac{\rho(k_0)}{(k_0-E_\phi)} \,dk_0 \nonumber \\
       \int^{\infty}_{-\infty}  \frac{{\rho(k_0)} }{(k_0-E_\phi)^2} \Big[1- {\cos\big((k_0-E_\phi)\,t\big)}    \Big]\,dk_0     & ~{}_{\overrightarrow{t\rightarrow \infty}} ~  & \pi \,t \Big[ \rho(E_{\phi}) + \mathcal{O}(1/t)\Big] \,,\label{asytos} \eea therefore asymptotically at long time
       \be \mathcal{E}(t) ~~ {}_{\overrightarrow{t\rightarrow \infty}} ~~ \Delta\,t -i \frac{\Gamma}{2}\, t\,,  \ee in agreement with the result (\ref{cfisol}).

  \section{Unitarity relation at all time:}\label{app:alltime}
With the results for $T_{\phi\gamma\gamma}$   and $C_{\gamma\gamma}(\vk_1,\lambda_1;\vk_2,\lambda_2)$ given by equations (\ref{Tfig},\ref{cgg}) respectively, it follows
that
\be \frac{1}{2!} \sum_{\vk_1,\lambda_1}\sum_{\vk_2,\lambda_2} |C_{\gamma\gamma}(\vk_1,\lambda_1;\vk_2,\lambda_2;t)|^2 = \sum_{\vk_1,\lambda_1}\sum_{\vk_2,\lambda_2}\frac{1}{2!}\, |T_{\phi \gamma\gamma}|^2\, \Bigg[\frac{1+e^{-\Gamma t} -\Big(e^{-i(E^r_{\phi}-k_1-k_2)t} + e^{i(E^r_{\phi}-k_1-k_2)t}\Big)\,e^{-\frac{\Gamma}{2}t}}{(k_1+k_2-E^r_{\phi})^2+\frac{\Gamma^2}{4}}\Bigg] \,,\label{ampsq}  \ee which can be written in terms of the spectral density (\ref{rhoax}) as
\be \int^\infty_{-\infty} dk_0 \, \rho(k_0) \,\Bigg[\frac{1+e^{-\Gamma t} -\Big(e^{i(k_0 -E^r_{\phi})t} + e^{-i(k_0-E^r_{\phi})t}\Big)\,e^{-\frac{\Gamma}{2}t}}{\big[(k_0-E^r_{\phi})-i\frac{\Gamma}{2}\big]\big[(k_0-E^r_{\phi})+i\frac{\Gamma}{2} \big]}\Bigg]\,.\label{ampsqrho}\ee   The integrand is dominated by the complex poles at $k_0= E^r_\phi \pm i \frac{\Gamma}{2}$, therefore to leading order in the coupling we replace $\rho(k_0) \rightarrow \rho(E_\phi)+\mathcal{O}(H^4_I)$ and carry out the remaining integral by residues in the complex $k_0$ plane with the final leading order result
\be \frac{1}{2!} \sum_{\vk_1,\lambda_1}\sum_{\vk_2,\lambda_2} |C_{\gamma\gamma}(\vk_1,\lambda_1;\vk_2,\lambda_2;t)|^2 = \frac{2\pi \rho(E_{\phi})}{\Gamma} \, \Big(1-e^{-\Gamma t}\Big) \equiv  \Big(1-e^{-\Gamma t}\Big) \,,\label{finiam}\ee where we used the relation (\ref{Gamafi}) and neglected contributions of $\mathcal{O}(H^2_I)$ and higher. Combined with the result (\ref{cifitsol}) for the amplitude $C_{\phi}(t)$ confirms the unitarity relation (\ref{unitycs}) to leading order in the interaction.

\section{Correlations}\label{app:corres}

\bea \langle 1_{\vk_1',\lambda_1'};1_{\vk_2',\lambda_2'}| a^\dagger_{\alpha'}({\vp}^{\,'}) \,a_{\alpha}(\vp)    |  1_{\vk_1,\lambda_1};1_{\vk_2,\lambda_2}\rangle & = &  \delta_{{\vp}^{\,'} \vk_1'}\,\delta_{\alpha' \lambda_1'}\delta_{\vp\, \vk_1} \delta_{\alpha \lambda_1} \delta_{\vk_2'\vk_2} \delta_{\lambda_2'\lambda_2} ~~~{(a)} \nonumber \\
& + & \delta_{{\vp}^{\,'} \vk_1'}\,\delta_{\alpha' \lambda_1'}\delta_{\vp\, \vk_2} \delta_{\alpha \lambda_2} \delta_{\vk_2'\vk_1} \delta_{\lambda_2'\lambda_1} ~~~{(b)} \nonumber\\
& + & \delta_{{\vp}^{\,'} \vk_2'}\,\delta_{\alpha' \lambda_2'}\delta_{\vp\, \vk_1} \delta_{\alpha \lambda_1} \delta_{\vk_1'\vk_2} \delta_{\lambda_1'\lambda_2} ~~~(c) \nonumber \\
& + & \delta_{{\vp}^{\,'} \vk_2'}\,\delta_{\alpha' \lambda_2'}\delta_{\vp\, \vk_2} \delta_{\alpha \lambda_2} \delta_{\vk_1'\vk_1} \delta_{\lambda_1'\lambda_1} ~~~{(d)}\label{adagera}
\eea  with $\vk'_2= \vk-\vk'_1~~;~~\vk_2=\vk-\vk_1$. The constraints yield for $(a)-(d)$
\bea (a): &&  \vk'_2=\vk_2; \vk'_1=\vk_1;\vp=\vp^{\,'}=\vk_1;\lambda'_1=\alpha';\lambda_1=\alpha;  \lambda'_2=\lambda_2 ~~(\mathrm{summed~over}) \label{consta}\\
(b): &&  \vk'_2=\vk_1; \vk'_1=\vk_2;\vp=\vp^{\,'}=\vk_2;\lambda'_1=\alpha';\lambda_2=\alpha;  \lambda'_2=\lambda_1 ~~(\mathrm{summed~over}) \label{constb}\\
(c): &&  \vk'_2=\vk_1; \vk'_1=\vk_2;\vp=\vp^{\,'}=\vk_1;\lambda'_2=\alpha';\lambda_1=\alpha;  \lambda'_1=\lambda_2 ~~(\mathrm{summed~over}) \label{constc}\\
(d): &&  \vk'_2=\vk_2; \vk'_1=\vk_1;\vp=\vp^{\,'}=\vk_2;\lambda'_2=\alpha';\lambda_2=\alpha;  \lambda'_1=\lambda_1 ~~(\mathrm{summed~over}) \label{constd}
\eea

\bea & & a_{s_3}(\vp_1) a_{s_4}(\vp_2) |  1_{\vk_1,\lambda_1};1_{\vk_2,\lambda_2}\rangle   = \nonumber\\
   & & \big( \delta_{\vp_1 \vk_1} \delta_{\vp_2 \vk_2}\delta_{s_3 \lambda_1}\delta_{s_4 \lambda_2} + \delta_{\vp_1 \vk_2} \delta_{\vp_2 \vk_1}\delta_{s_3 \lambda_2}
   \delta_{s_4 \lambda_1} \big)\ket{0;0} \nonumber \\
   & & \langle 1_{\vk_1',\lambda_1'};1_{\vk_2',\lambda_2'}|a^{\dagger}_{s_1}(\vp_1) a^{\dagger}_{s_2}(\vp_2) = \nonumber \\ & &  \big(\delta_{\vp_1 \vk_1'} \delta_{\vp_2 \vk_2'}\delta_{s_1 \lambda_1'}\delta_{s_2 \lambda_2'}+ \delta_{\vp_1 \vk_2'} \delta_{\vp_2 \vk_1'}\delta_{s_1 \lambda_2'}\delta_{s_2 \lambda_1'}  \big)\bra{0;0}\,,\label{foura}
 \eea


\begin{thebibliography}{99}

\bibitem{PQ} R. D. Peccei and H. R. Quinn,  {CP Conservation in the Presence of Pseudoparticles}, Phys. Rev. Lett. 38, 1440
(1977), {Constraints imposed by CP conservation in the presence of pseudoparticles}, Phys. Rev. D 16, 1791 (1977).

\bibitem{weinaxion} S. Weinberg,   {A New Light Boson?}, Phys. Rev. Lett. 40,
223 (1978).

\bibitem{wil} F. Wilczek,   {Problem of Strong P and T Invariance in the Presence of Instantons}, Phys. Rev. Lett. 40, 279 (1978).




 \bibitem{marsh} D.J.E. Marsh,  {Axion Cosmology },  Phys. Rept., 643, 1  (2016);  F. Chadha-Day, J. Ellis, D. J. E. Marsh, {Axion Dark Matter: What is it and Why Now?}, Sci. Adv. 8 eabj3618 (2022).

\bibitem{sikivie1} P. Sikivie,  {Invisible axion search methods}, Rev. Mod. Phys. 93, 015004 (2021).


\bibitem{press} J. Preskill, M. B. Wise, F. Wilczek, {Cosmology of the Invisible Axion} Phys.Lett. 120B, 127  (1983).

\bibitem{abbott} L.F. Abbott, P. Sikivie, {A Cosmological Bound on the Invisible Axion }  Phys.Lett. B120 133 (1983).

\bibitem{fischler} M. Dine, W.  Fischler, {The Not So Harmless Axion},  Phys.Lett. B120, 137  (1983).


\bibitem{wilczekaxion} F. Wilczek,  {Two applications of axion electrodynamics}, Phys. Rev. Lett. 58, 1799 (1987).




\bibitem{xiao} X.-Liang Qi, T. L. Hughes, S.-Cheng Zhang,  {Topological field theory of time-reversal invariant insulators}, Phys. Rev. B78, 195424 (2008).





     \bibitem{rundong} R. Li, J. Wang, X-L Qi, S-C. Zhang,  {Dynamical axion field in topological magnetic insulators},  Nature Physics 6, 284 (2010).

     \bibitem{jing} J. Wang, B. Lian, S-C. Zhang,  {Dynamical axion field in a magnetic topological insulator superlattice}, Phys. Rev B 93, 045115 (2016). 

 \bibitem{nomura} A. Sekine, K. Nomura,  {Axion electrodynamics in topological materials}, Jour. of Applied Physics, 129, 141101 (2021).

 \bibitem{narang} D. M. Nenno, C. A. C. Garcia, J. Gooth, C. Felser, P. Narang,  {Axion physics in condensed-matter systems},  Nature Reviews Physics 2, 682 (2020).



 \bibitem{gooth} J. Gooth, B. Bradlyn, S. Honnali, C. Schindler, N. Kumar, J. Noky, Y. Qi, C. Shekhar, Y. Sun, Z. Wang, B. A. Bernevig, C. Felser,  {Axionic charge-density wave in the Weyl semimetal (TaSe4)2I},  Nature 575, 315 (2019).

   \bibitem{gos}   Pallab Goswami, Sumanta Tewari, { Axionic field theory of (3+1)-dimensional Weyl semi-metals}, Phys. Rev. B 88, 245107 (2013).

 \bibitem{yu} J. Yu,B. J. Wieder, C.-X. Liu,  {Dynamical piezomagnetic effect in time-reversal-invariant Weyl semimetals with axionic charge density waves}, Phys. Rev. B104, 174406 (2021).



  \bibitem{mottola}  E. Mottola, A. V. Sadofyev, A. Stergiou,  {Axions and superfluidity in Weyl semimetals}, Phys. Rev. B 109, 134512 (2024).


 \bibitem{wilczekshapo} L. Shaposhnikov, M. Mazanov, D. A. Bobylev, F. Wilczek, M. A. Gorlach,  {Emergent axion response in multilayered metamaterials}, Phys. Rev. B 108, 115101 (2023).

 \bibitem{liang} L. Wu, M. Salehi, N. Koirala, J. Moon, S. Oh, N. P. Armitage, {Quantized Faraday and Kerr rotation and axion electrodynamics of a 3D topological insulator}, Science 354, 1124 (2016).

 \bibitem{tse} W. K. Tse, A. H. MacDonald,  {Giant Magneto-Optical Kerr Effect and Universal Faraday Effect in Thin-Film Topological Insulators}, Phys. Rev. Lett. 105, 057401 (2010).

 \bibitem{ahn} J. Ahn, S.-Y. Xu, A. Vishwanath,  {Theory of optical axion electrodynamics and application to the Kerr effect in topological antiferromagnets}, Nature Communications 13, 7615 (2022).


 \bibitem{ishi} K. Ishiwata, K. Nomura,  {Collective excitations in magnetic topological insulators and axion dark matter search},  JHEP10 (2024) 225.

 \bibitem{jan} J. Schütte-Engel, D. J. E. Marsh, A. J. Millar, A. Sekine, F. Chadha-Day, S. Hoof, M. Ali, K.-C. Fong, E. Hardy, L. Smejkal, {Axion Quasiparticles for Axion Dark Matter Detection}, JCAP 08, 066 (2021).

 \bibitem{zhu} T. Zhu, H. Wang1,  D. Xing,  H. Zhang, Axionic surface wave in dynamical axion insulators,  Phys. Rev. B 106, 075103 (2022).

 \bibitem{marsh2} D. J. Marsh, K. C. Fong, E. W.  Lentz, L. Smejkal, M. N.  Ali, {Proposal to detect dark matter
using axionic topological antiferromagnets.},  Phys. Rev. Lett. 123, 121601 (2019).

\bibitem{chig} S. Chigusa, T.  Moroi,  K.  Nakayama, {Axion-hidden-photon dark matter conversion into
condensed matter axion.} J. High Energy Phys. 2021, 1 (2021).

 \bibitem{boyhybrid} D. Boyanovsky, {Synthetic and cosmological axion hybridization:  entangled photons, (HBT) and quantum beats.}, Phys. Rev. Research 7, 033164 (2025).

 \bibitem{zaletel} Z.-Q. Gao, T. Wang, M. P. Zaletel, D.-H. Lee,  {Detecting axion dynamics on the surface of magnetic topological insulators}, Phys. Rev. B 111, 214407 (2025).



 \bibitem{jianq} Jian-Xian Qiu, \emph{et.al.} {Observation of the axion quasiparticle in 2D
MnBi2Te4}, Nature 641, 62 (2025).




 \bibitem{ww} V. Weisskopf, E. Wigner,  {Calculation of the natural brightness of spectral lines on the basis of Dirac's theory},  Z. Phys. 63, 54 (1930).

 \bibitem{zubairy} M. O. Scully, M. S. Zubairy,  {Quantum Optics} (Cambridge University Press, Cambridge, UK, 1997).


\bibitem{meystre} P. Meystre, M. Sargent III,  {Elements of Quantum Optics, 4th Ed.} (Springer, N.Y. 2007).


 \bibitem{born} M. Born, E. Wolf, \emph{Principles of optics} (Cambridge University Press, Cambridge, UK, 1999).

 \bibitem{mandl} L. Mandel, E. Wolf, \emph{Optical Coherence and Quantum Optics}, (Cambridge University Press, Cambridge U.K.1995).

 \bibitem{jauch} J. M. Jauch, F. Rohrlich, \emph{The theory of Photons and Electrons} (Addison-Wesley, Reading, Massachussetts, 1955).

 \bibitem{agarwal} G. S. Agarwal, \emph{Quantum Optics} (Cambridge University Press, Cambridge, UK, 2013).

\bibitem{stokes1} G. Bjork, J. Soderholm, L. L. Sanchez-Soto, A. B. Klimov, J. Ghiu, P. Marian, T. A. Marian,  {Quantum degrees of polarization}, Optics Communications 283, 4440 (2010).

    \bibitem{qupola1} A. Z. Goldberg, P. de la Hoz, G. Bjork, A. B. Klimov, M. Grassi,  G. Leuchs, L. Sanchez-Soto, {Quantum concepts in optical polarization}, Advances in Optics and Photonics, Vol 15, No.1, 1  (2021).

    \bibitem{aga} G.S. Agarwal, S. Chaturvedi, {Scheme to measure quantum Stokes parameters and their fluctuations and correlations},  Jour. of Mod. Optics 50, 711 (2002).


 \bibitem{korolkova} N. Korolkova, G. Leuchs, R. Loudon, T. C. Ralph, C. Silberhorn, {Polarization squeezing and continuous-variable polarization entanglement}, Phys. Rev. A 65, 052306 (2002).




























\bibitem{glauber} R. J. Glauber,  {Optical coherence and photon statistics} in  Quantum Optics and
Electronics (eds DeWitt, C. Blandin, A.  Cohen-Tannoudji, C.) 63 (Gordon and
Breach, New York , 1965);  {Photon Correlations}, Phys. Rev. Lett. 10, 84 (1963).



\bibitem{hbt} R. Hanbury Brown,  R. Q.  Twiss,   {Correlation between photons in two coherent beams of light},  Nature 177, 27  (1956);  {A test of a new type of stellar interferometer on Sirius}, Nature 178, 1046 (1956);  {Interferometry of the intensity fluctuations in light-I. Basic theory: the correlation between photons in coherent beams of radiation}, Proc. Roy. Soc. London, Series A, Vol 242, 300 (1957).

    \bibitem{hbtcolor}     B. Silva, C. Sánchez Muñoz, D. Ballarini, A. González-Tudela, M. de Giorgi, G. Gigli, K. West, L. Pfeiffer, E. del Valle, D. Sanvitto, F. P. Laussy,  {The colored Hanbury Brown–Twiss effect}, Scientific Reports 6, 37980 (2016).








\bibitem{pdc1} R. Ghosh, C. K. Hong, Z. Y. Ou, L. Mandel,  {Interference of two photons in parametric down conversion}, Phys. Rev. A 34, 3962 (1986).

\bibitem{pdc2} C.K. Hong, Z. Y. Ou, L. Mandel,  {Measurement of subpicosecond time intervals between two photons by interference}, Phys. Rev. Lett. 59, 2044 (1987).

\bibitem{pdc3} M. H. Rubin, D. N. Klyshko, Y. H. Shih, A. V. Sergienko,  {Theory of two-photon entanglement in type-II optical parametric down-conversion}, Phys. Rev. A 5122 (1994).


\bibitem{bi} Z. Y. Ou, L. Mandel,  {Violation of Bell's Inequality and Classical Probability in a Two-Photon Correlation Experiment}, Phys. Rev. Lett. 61, 50 (1988).

\bibitem{kiess} T. E. Kiess, Y. H. Shih, A. V. Sergienko, C. O. Alley, {Einstein-Podolsky-Rosen-Bohm experiment using pairs of light quanta produced by type-II parametric down-conversion}, Phys. Rev. Lett. 71, 3893 (1993).


\bibitem{shih} Y. Shih,  C. Alley, New type of Einstein-Podolsky-Rosen-Bohm experiment using pairs of light quanta pro-
duced by optical parametric down conversion, Physical
Review Letters 61, 2921 (1988).



\bibitem{exci}     R. M. Stevenson, R. J. Young, P. Atkinson, K. Cooper, D. A. Ritchie,  A. J. Shields,  {A semiconductor source of triggered entangled photon pairs}, Nature 439, 179 (2006).  

 \bibitem{tur} Noam Tur, Ismail Nassar, Ido Schwartz, Joseph Avron, Dan Dalacu, Philip J. Poole, David Gershoni, {Temporal CW polarization-tomography of photon pairs from the biexciton radiative cascade: theory and experiment}, Phys. Rev. B 111, 235304 (2025).


    \bibitem{dousse} A. Dousse, J. Suffczynski, A. Beveratos, O. Krebs, A. Lemaıtre, I. Sagnes,
J. Bloch, P. Voisin,  P. Senellart, {Ultrabright source of entangled photon pairs}, Nature Letters, 466, 217 (2010). 

\bibitem{akopian} N. Akopian, N. H. Lindner, E. Poem, Y. Berlatzky, J. Avron, D. Gershoni, B. D. Gerardot, P. M. Petroff
 {Entangled photon pairs from semiconductor quantum dots}, Phys. Rev. Lett. 96, 130501 (2006).









\bibitem{kwiat1} P. Kwiat, K. Mattle, H. Weinfurter, A. Zeilinger,
A. Sergienko, and Y. Shih, New high-intensity source
of polarization-entangled photon pairs, Physical Review
Letters 75, 4337 (1995).


\bibitem{kwiat2} P. Kwiat, E. Waks, A. White, I. Appelbaum, P. Eberhard, Ultrabright source of polarization-entangled pho-
tons, Physical Review A 60, R773 (1999).

    \bibitem{delgado} A. J. Delgado, G. Giedke, J. Aizpurua, R. Esteban, Generation of polarization-entangled photon pairs from two interacting quantum emitters, arXiv: 2503.02739.

\bibitem{olivia} O. Liebman, J. Curtis, I. Petrides, P. Narang,  {Multiphoton Spectroscopy of a Dynamical Axion Insulator}, arXiv:2306.00064  









\bibitem{ouwang} Z. Y. Ou, J.-K. Rhee, J. J. Wang, {Photon bunching and multiphoton interference in parametric down conversion.}, Phys. Rev. A 593 (1999).



\bibitem{stokes2} D. Kuebel, T. D. Visser,  {Generalized Hanbury Brown-Twiss effect for Stokes parameters}, Journal of the Optical Society of America, A36, 362 (2019).

\bibitem{stokes3} X. Liu, G. Wu, X. Pang, D. Kuebel, T. D. Visser,  {Polarization and coherence in the Hanbury Brown-Twiss effect}, Journal of Mod. Optics 65, 1437 (2018).



\end{thebibliography}
  \end{document}